\PassOptionsToPackage{table}{xcolor}

\documentclass[twocolumn]{aastex631}
\usepackage{showyourwork}
\usepackage{amsmath, amssymb}
\usepackage{gensymb}
\usepackage{xcolor}
\usepackage{booktabs}
\usepackage{dsfont}
\usepackage[inline]{enumitem}
\usepackage{cleveref}
\usepackage{transparent}
\usepackage{array}
\usepackage{float}

\newcommand{\code}[1]{\textsc{#1}}
\newcommand{\package}[1]{\code{#1}}
\newcommand{\stream}[1]{#1}
\newcommand{\dataarchive}[1]{\textit{#1}}
\newcommand{\Gaia}{\dataarchive{Gaia}}
\newcommand{\PanStarrs}{\dataarchive{Pan-STARRS1}}

\newcommand{\mrm}[1]{\mathrm{#1}}
\newcommand{\mbs}[1]{\boldsymbol{#1}}
\newcommand{\mbf}[1]{\mathbf{#1}}

\newcommand{\mcal}[1]{\mathcal{#1}}

\newcommand{\Exp}[1]{e^{#1}}

\NewDocumentCommand \dif {o m} {
	\IfNoValueTF{#1}
		{ \mathrm{d}{#2} }
		{ \mathrm{d}^{#1}{\!#2} }
}

\newcommand{\pdf}{p}
\newcommand{\cdf}{P}

\newcommand{\prior}{\mcal{\pi}}

\newcommand{\nth}[1]{{#1}_{\mrm{n}}}  
\newcommand{\fth}[1]{{#1}_{\mrm{f}}}  
\newcommand{\qth}[1]{{#1}_{\mrm{q}}}  
\newcommand{\unit}[1]{[\text{#1}]}

\newcommand{\smallcomponent}[2]{#2^{\scriptscriptstyle (#1)}}
\newcommand{\cmp}[2]{\smallcomponent{#1}{#2}}
\newcommand{\Scmp}[1]{\cmp{S}{#1}}
\newcommand{\Bcmp}[1]{\cmp{B}{#1}}
\newcommand{\astroM}[1]{{#1}_w}

\newcommand{\Spdf}{\Scmp{\pdf}}
\newcommand{\Bpdf}{\Bcmp{\pdf}}

\newcommand{\parallax}{\varpi}
\newcommand{\sigobs}{{\sigma_*}}

\defcitealias{Ibata+2021}{I21}

\begin{document}

\title{Stream Members Only:\\ Data-Driven Characterization of Stellar Streams with Mixture Density Networks}

\author[0000-0003-3954-3291]{Nathaniel Starkman}
\altaffiliation{Co-first author.}
\affiliation{David A Dunlap Department of Astronomy and Astrophysics, University of Toronto, 50 St. George St, Toronto ON M5S 3H4, Canada}

\author[0000-0001-8042-5794]{Jacob Nibauer}
\altaffiliation{Co-first author.}
\affiliation{Department of Astrophysical Sciences, Princeton University, 4 Ivy Ln, Princeton, NJ 08544, USA}

\author[0000-0001-6855-442X]{Jo Bovy}
\affiliation{David A Dunlap Department of Astronomy and Astrophysics, University of Toronto, 50 St. George St, Toronto ON M5S 3H4, Canada}

\author[0000-0003-3613-0854]{Jeremy J. Webb}
\affiliation{Department of Science, Technology and Society, York University, 4700 Keele St, Toronto ON M3J 1P3, Canada}

\author[0000-0001-6584-6144]{Kiyan Tavangar}
\affiliation{Department of Astronomy, Columbia University, 538 West 120th Street, New York, NY 10027, USA}

\author[0000-0003-0872-7098]{Adrian Price-Whelan}
\affiliation{Center for Computational Astrophysics, Flatiron Institute, 162 Fifth Ave, New York, NY 10010, USA}

\author[0000-0002-7846-9787]{Ana Bonaca}
\affiliation{The Observatories of the Carnegie Institution for Science, 813 Santa Barbara Street, Pasadena, CA 91101, USA}

\correspondingauthor{Nathaniel Starkman}
\email{n.starkman@mail.utoronto.ca}

\shortauthors{Starkman and Nibauer et al.}

\begin{abstract}

    Stellar streams are sensitive probes of the Milky Way's gravitational
    potential. The mean track of a stream constrains global properties of the
    potential, while its fine-grained surface density constrains galactic
    substructure. A precise characterization of streams from potentially noisy
    data marks a crucial step in inferring galactic structure, including the
    dark matter, across orders of magnitude in mass scales. Here we present a
    new method for constructing a smooth probability density model of stellar
    streams using all of the available astrometric and photometric data. To
    characterize a stream's morphology and kinematics, we utilize mixture
    density networks to represent its on-sky track, width, stellar number
    density, and kinematic distribution. We model the photometry for each stream
    as a single-stellar population, with a distance track that is simultaneously
    estimated from the stream's inferred distance modulus (using photometry) and
    parallax distribution (using astrometry). We use normalizing flows to
    characterize the distribution of background stars. We apply the method to
    the stream \stream{GD-1}, and the tidal tails of \stream{Palomar 5}. For
    both streams we obtain a catalog of stellar membership probabilities that
    are made publicly available.  Importantly, our model is capable of handling
    data with incomplete phase-space observations, making our method applicable
    to the growing census of Milky Way stellar streams. When applied to a
    population of streams, the resulting membership probabilities from our model
    form the required input to infer the Milky Way's dark matter distribution
    from the scale of the stellar halo down to subhalos.

\end{abstract}

\section{Introduction} \label{sec:intro}

    Stellar streams are the disrupted remnants of globular clusters and
    satellite galaxies, and their census has grown extensively in the Milky Way
    thanks to all-sky astrometric missions like \Gaia{}
    \citep{GaiaCollaboration+2016, GaiaCollaboration+2023} and low-surface
    brightness photometric surveys (e.g., \citealt{Ibata+1998,
    GrillmairDionatos2006, Shipp+2018, Ibata+2021}).  The formation of streams
    can be traced back to a progenitor (a globular cluster or satellite galaxy)
    whose stars are gradually lost to the tidal field of the host galaxy. The
    stars extend along a series of similar orbits, called tidal tails, which can
    sometimes span several tens of degrees across the sky. The orbital proximity
    of stars belonging to a stream makes them sensitive tracers of the mass
    distribution of galaxies, sourced by both the baryonic and dark matter
    components.

    The average properties of a stream (e.g., the mean position and velocity
    track of the tidal tails) can be used to reconstruct the global dark matter
    distribution of the stellar halo (e.g., \citealt{Johnston+1999, Bonaca+2014,
    Bovy2014, Bovy+2016, Nibauer+2022, Koposov+2023}), while local variations in
    the stream track and surface density encode information about the stream's
    progenitor \citep{Kupper+2012, Sanders+2013, Price-Whelan+2014}, encounters
    with baryonic structures such as the stellar bar (e.g., \citealt{Erkal+2017,
    Pearson+2017}), and non-luminous structures like dark matter subhalos (e.g.,
    \citealt{Ibata+2002, Johnston+2002, Carlberg+2012-74820C, Bovy+2017,
    Bonaca+2019, Hermans+2021}).

    While several methods have been developed to model stream properties as a
    function of global halo characteristics (e.g., the mass distribution,
    flattening; \citealt{Johnston+1999, Binney2008, Koposov+2010,
    SandersBinney2013, Bovy2014, Nibauer+2022}) and local substructures (e.g.,
    the bar's pattern speed and the mass of subhalo perturbers;
    \citealt{Antoja+2014, Hattori+2016, Price-Whelan+2016, Bovy+2017,
    Pearson+2017, Bonaca+2019}), these methods often rely on the existence of a
    homogenized stream-tracer population that characterizes a given stream
    across each of the observable phase-space dimensions.  Additionally, because
    even the fine-grained properties of streams are influenced by the detailed
    mass distribution of its host galaxy, it is important to propagate errors
    both in measurement uncertainty and membership uncertainty (i.e., which
    stars belong to the stream versus the background) when attempting to model a
    stream. Otherwise, a poorly characterized stream could bias constraints on,
    e.g., the total mass of the galaxy, the dark matter halo shape, and the mass
    of subhalo perturbers.  In order to generate more precise constraints on the
    structure of the galaxy, it is therefore imperative that a statistically
    sound and homogenized catalog of stellar streams is produced. 

    While streams are well described in action-angle coordinates
    \citep[e.g.][]{Bovy2014} or (restricted) $N$-body simulations
    \citep[e.g.][]{Dehnen+2004}, these methods rely on prior assumptions about
    the galaxy and its underlying gravitational potential.  Therefore, it is
    necessary to devise data-driven methods to characterize stellar streams
    without appealing to strong prior assumptions about the galaxy's structure. 
    
    There are a few previous works which have modeled stellar streams in this
    context. \citet{Patrick+2022} developed the method introduced by
    \citet{Erkal+2017}, which is a data-driven, spline-based method for modeling
    the photometric properties and on-sky positions of previously identified
    streams. Their method works by fitting the stream's stellar surface density
    with a series of splines, capturing both surface brightness variations along
    the stream and changes in the position of the stream track and width. Their
    method also utilizes an isochrone fit to a given stream, from which a
    distance modulus can be derived. While this method is very successful at
    characterizing the photometric properties of a stream, it does not consider
    the kinematic dimensions which have become increasingly well-measured with
    successive data releases from \Gaia{}, as well as targeted observations.
    From a population of RR Lyrae stars, \citet{Price-Whelan+2019} utilized a
    mixture model to fit the globular cluster stream Pal-5 in the space of
    proper motions and heliocentric distances.  The mixture modeling approach
    enables a simultaneous fit to both the stream and background density,
    allowing for a cleaner selection of stream stars. However, the track of the
    stream in on-sky coordinates is not fit. The method identified 27 RR Lyrae
    stars consistent with being members of the stream. While useful for
    determining the average properties of a stream (e.g., its mean distance
    track), a limited population of RR Lyrae stars are not sufficient to capture
    the small-scale properties of streams expected to inform constraints on,
    e.g., a population of perturbers in the galaxy.

    Other methods for extracting stellar streams from the data include
    matched-filter algorithms, where over-densities in color-magnitude space are
    inspected as possible clusters undergoing tidal stripping
    \citep{Grillmair+1995, GrillmairJohnson2006, Shipp+2018}.  While the
    matched-filter technique has been extremely successful in stream discovery,
    it is not equipped to characterize streams at the level of detail necessary
    to inform precise constraints on the potential of the galaxy and its
    substructure.

    In order to address the need for a flexible model of stellar streams both in
    photometric and kinematic spaces, in this work we develop a new method for
    characterizing streams and quantifying the membership probability of
    possible stream stars. Our approach provides several key advantage over
    previous stream modeling efforts, specifically in its ability to jointly
    model the kinematic and photometric properties of streams simultaneously
    using a flexible model.  Furthermore, our method is well-suited to fitting
    streams with only partial or incomplete phase-space or photometric
    observations, enabling a full characterization of streams insofar as the
    available data allows.

    \paragraph{Paper organization}

        The paper is organized as follows.  In \autoref{sec:method} we introduce
        our method for characterizing stream tracks and computing stellar
        membership probabilities.  We discuss initial data processing in
        \autoref{sub:method:framing_the_data} and build a probabilistic
        framework to model the stream and background in
        \autoref{sub:method:likelihood_setup}-\autoref{sub:methods:priors}.  In
        \autoref{sec:results_mock} we apply our method to simulated data to
        demonstrate the model with a known ground truth dataset, then to the
        stellar streams \stream{GD-1} (\autoref{sec:results_gd1}) and
        \stream{Pal\,5} (\autoref{sec:results_pal5}).  We compare our model to
        other methods in \autoref{sec:discussion}.  We conclude in
        \autoref{sec:conclusions}, discussing the results of this work and
        future directions.


\newpage
\section{Method} \label{sec:method}

    Below, we offer a concise overview of our approach, followed by a more
    in-depth technical explanation in subsequent sections.  We start, in
    \autoref{sub:method:framing_the_data}, by transforming the astrometric data
    into a stream-oriented on-sky reference frame ($\phi_1, \phi_2, \parallax,
    \mu_{\phi_1}^*, \mu_{\phi_2}, v_r)$, for each stream.  In this frame, the
    stream track lies along $\phi_1$, and the stream density may be described by
    distributions in the other parameters as injective functions of $\phi_1$
    (called conditional distributions).  Likewise, the background of the
    Galactic field may be described with $\phi_1$-conditioned distributions.  We
    will also model the stream in photometric coordinates.  Additional features,
    e.g.  metallicity, can also be modeled along the stream and background as
    functions of $\phi_1$, though we leave this to a future work.

    The distributions for the stream and background are collected in a single
    mixture model that describes the entire field.  We take the parameters of
    the mixture model --- the mixture weights and component distribution
    parameters --- to be general (continuous) functions of $\phi_1$, the
    conditioning variable. This is done using a feed-forward neural network that
    takes $\phi_1$, the conditioning variable, as its input and outputs the
    parameter value(s) that characterize the distribution of stars at a $\phi_1$
    ``slice".  Mixture models of conditional probability distributions using
    neural networks are called Mixture Density Networks (MDNs)
    \citep{Bishop+1994}.  Since the model coefficients may evolve with $\phi_1$,
    MNDs are well suited to characterize streams (and the background) with all
    their variations in position, width, and linear density.


    \subsection{Framing the Data} \label{sub:method:framing_the_data}

        As a stream progenitor orbits the Galactic center of mass, its path can
        be approximated locally as an elliptical segment. The associated arms of
        the stream generally align closely with its orbital plane.  As viewed
        from a Galactocentric reference frame, the stream approximately assumes
        the form of a great circle enabling a 1:1 mapping from some phase
        parameter to a unique position along the stream. Thus in a `stream'
        oriented frame the rotated longitudinal and latitudinal coordinates
        ($\phi_1, \phi_2$) align with the stream such that the stream's phase is
        $\phi_1$ and $\phi_2 \simeq 0$.

        While it would be ideal to characterize streams in this Galactocentric
        frame, transformation to the frame requires knowledge of the distance to
        the Galactic center  \citep{GRAVITY2018, BennettBovy2019, Leung+2022},
        the solar motion around the Galactic center, and accurate distances to
        field stars (among other uncertain quantities). In a heliocentric frame,
        e.g. the standard ICRS\footnote{International Celestial Reference
        System} \citep{ICRS1997}, projection effects can obscure the shape of a
        stream leading to a non-trivial stream track. Therefore in a
        heliocentric frame, there is no assurance of effecting a rotation like
        in the Galactocentric frame that would similarly align the stream in an
        observer's coordinates such that $\phi_2(\phi_1) \approx 0$.
        Thankfully, most streams are adequately distant from both the Galactic
        center and the observer, or exhibit orientations such that a rotation a)
        can be established or b) holds valid. For most of the best-studied
        globular cluster streams -- \stream{GD-1}, \stream{Palomar 5},
        \stream{Jhelum}, etc. -- this holds true. However for some streams of
        dwarf galaxies, like the Magellanic and Sagittarius streams which each
        wrap around the Milky Way \citep{Newberg+2002, Majewski+2003,
        WannierWrixon1972}, no such transformation is possible. More generally,
        a stream's ``sky"-projected path may be kinked, e.g. by subhalo
        interactions, such that it is a many-to-one function in $\phi_1$.

        In this work we consider the vast majority of streams that can be
        described as an injective function, $\phi_2(\phi_1)$.  Since we are
        examining known streams this transformation is known \textit{a priori},
        or may be constructed during pre-processing
        \citep[see][\S2.1]{Starkman+2023}.  We note that, with sufficient prior
        knowledge, many-to-one functions may be broken into injective segments
        and their connection constrained by a prior. Thus the methods developed
        here may be extended and applied even to wrapped and kinked streams.
        However, we leave extending the model framework to a future work.

    \subsection{Likelihood Setup}\label{sub:method:likelihood_setup}

        \begin{figure*}
            \script{pgm.py}
            \centering
            \includegraphics[scale=0.6]{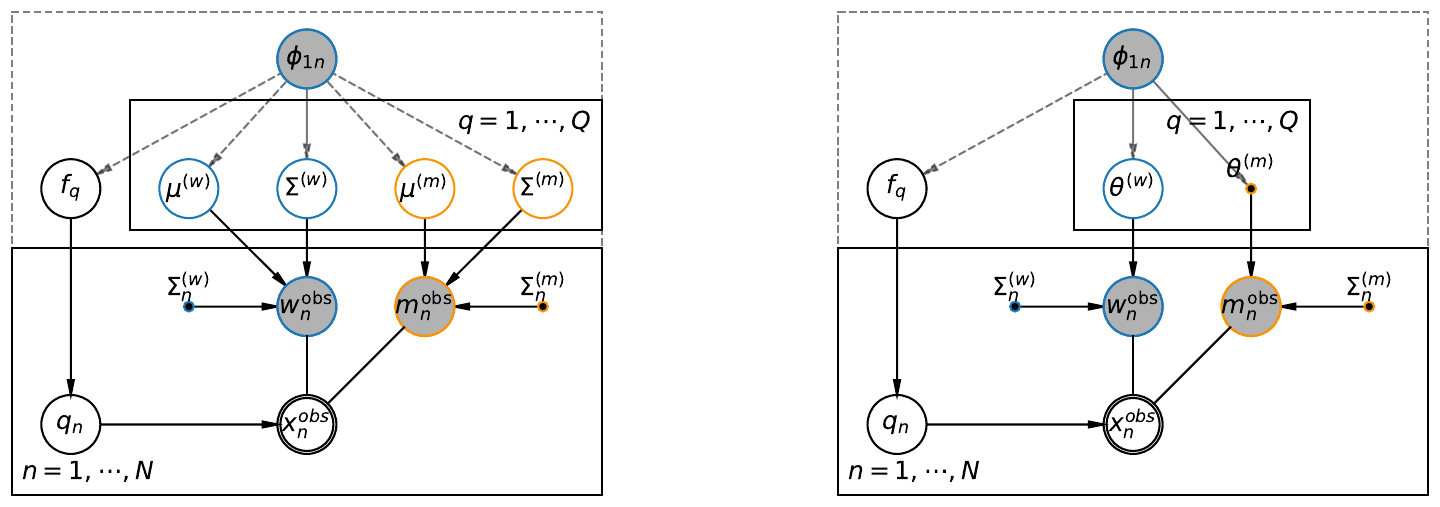}
            \caption{%
                Probabilistic Graphical Models (PGMs) of the Mixture Density
                Networks used in this work.
                \textbf{Left}: A PGM of the stream model. The full dataset
                $x_n^{obs}$ is the aggregation of the observed astrometry
                $w_n^{obs}$ and photometry $m_n^{obs}$, along with the
                corresponding observational errors. We model the data as a
                mixture of Gaussians (in astrometry + distance modulus), with
                the Gaussians indexed by $q$. The weight $\alpha$ determines the
                weighted contribution of each Gaussian to each datum. All models
                are conditioned on $\phi_1$.
                \textbf{Right}: a PGM of the background model. The data are
                identical to that of the stream, however the model is not a
                mixture of Gaussians, but of a variety of distributions. We
                further distinguish between trainable distribution parameters
                (circles) and fixed distributions (points), like a pre-trained
                normalizing flow background model, explained in further detail
                in \autoref{sub:method:pre-training_distributions}. Alike to the
                stream model, all background models are conditioned on $\phi_1$.
            }
            \label{fig:PGM}
        \end{figure*}

        Mixture models are a statistical method to represent a population as a
        set of a sub-populations specified by probability distributions.  These
        distributions are characterized by their parameters and the mixture by a
        corresponding set of mixture coefficients. Scalar mixture models may be
        extended to model completely general conditional distributions by making
        the distribution parameters and mixture coefficients into functions of a
        conditional input \citep{McLachlanBasford1989}.  Mixture Density
        Networks (MDNs) are a machine learning method to implement these general
        mixture models, taking the parameters and coefficients to be the outputs
        of feed-forward neural networks and as such, general continuous
        functions \citep{Bishop1994}.  Functional parameters allow the model to
        capture variation over those parameters that is not possible with
        scalar-valued models. Moreover, by using neural networks the inferred
        functional forms are ``model-free" and driven entirely by the data.

        For the MDN we take $x_{\cdot, 0}$, the first feature column of the data
        $\{\mbf{x}\}$, as the `independent' coordinate over which the MDN
        parameters are conditioned. In the context of streams $x_{\cdot, 0}
        \equiv \phi_1$ is the longitude coordinate in a reference frame rotated
        such that the stream is aligned along $\phi_1$. The parameters of the
        stream, the background field, and the mixing coefficients of the two are
        now general data-driven continuous functions along $\phi_1$.

        The PDF of a general MDN is
        \begin{equation} \label{eq:general_mixture_network}
            \!\!\! \pdf(\nth{\mbs{x}} | \mbs{\theta}(\phi_1))
            \!=\! \sum_{\mrm{q} \in \mbs{I_Q}} \! \qth{f}(\phi_1) \cmp{q}{\pdf}\left(\nth{\mbs{x}}|\qth{\mbs{\theta}}(\phi_1)\right),
        \end{equation}
        where $\cmp{q}{\pdf}$ is the PDF of the $q$-th model and $f_q$ its
        mixing coefficient for all $q$ models in the set of models indexed by
        $\mbs{I_Q}$. $\mbs{\theta}$ is the $\phi_1$-conditional output of the
        feed-forward neural network, and $\nth{\mbs{x}}$ the data on the $n$-th
        star in $\{\mbs{x}\}$, not including $\phi_1$. For convenience we drop
        the implied subscript $n$ in $\phi_1$, the explicit notation of
        $\mbs{\theta}$'s dependence on $\phi_1$, and the index set $\mbs{I}_X$
        in summations over its elements $x$.

        The mixture coefficients $\qth{f}$ are normalized such that
        \begin{equation} \label{eq:mixture_weight_sum}
            \sum_{\mrm{q}} \qth{f}(\phi_1) = 1, \qquad \forall \phi_1
        \end{equation}
        In practice this is enforced by defining a background
        \begin{equation} \label{eq:mixture_weight_normalization}
            f_{b}(\phi_1) = 1 - \sum_{\rm{q}\in \mbs{I_Q}\backslash \{b\}} \qth{f}(\phi_1),
        \end{equation}
        the summation over all models except for some ``background" index $b \in
        \mbs{I_Q}$.  In the context of Galactic streams the ``foreground" is the
        stream itself, while the background is the Galaxy --- bulge, bar, disc,
        halo --- and additional structures, like globular clusters and dwarf
        galaxies, that are not associated with the stream itself.  We discuss
        the mixture weights further in \autoref{sub:method:mixture_weight}.

        In this work we consider primarily astrometric and photometric data,
        (notated $\{\mbs{w}\}$ and $\{\mbs{m}\}$, respectively), but note that
        the MDN framework may be extended to arbitrary feature dimensions.
        Assuming the conditional independence of the astrometry and
        photometry\footnote{ For \Gaia{} data this is true in practice but not
        in detail: there exist small correlations between the astrometric and
        photometric data.}, we split each model into linearly separable
        astrometric and photometric models $\cmp{q}{\pdf} = \cmp{q,w}{\pdf}
        \cmp{q,m}{\pdf}$.

        The likelihood for a single star then becomes, simplifying from
        \autoref{eq:general_mixture_network},
        \begin{equation} \label{eq:general_model}
            \!\!\!\!\! \pdf(\nth{\mbs{w}}, \! \nth{\mbs{m}} | \mbs{\theta})
                \!=\! \sum_{q} \! f_q(\phi_1) \cmp{q,w}{\pdf}(\nth{\mbs{w}}|\mbs{\theta}) \cmp{q,m}{\pdf}(\nth{\mbs{m}}|\mbs{\theta})
        \end{equation}

        Assuming the measurement of each star is independent, the total
        log-likelihood is a sum over all stars:
        \begin{equation} \label{eq:general_likelihood}
            \ln\mcal{L}\left(\{\nth{\mbs{w}},\nth{\mbs{m}}\} | \mbs{\theta}\right) = \sum_n \ln \pdf(\nth{\mbs{w}}, \nth{\mbs{m}} | \mbs{\theta}).
        \end{equation}

        It will prove convenient to distinguish the models for $q$ as the
        \textit{background} and \textit{stream} models, respectively. In this
        notation:
        \begin{align} \label{eq:stream_and_bkg_prob}
            \pdf(\nth{\mbs{w}}, \nth{\mbs{m}} | \mbs{\theta})
            =& \phantom{+} \qquad f \phantom{+} \Spdf(\nth{\mbs{w}}|\mbs{\theta}) \Spdf(\nth{\mbs{m}}|\mbs{\theta}) \\
            & + (1-f) \Bpdf(\nth{\mbs{w}}|\mbs{\theta}) \Bpdf(\nth{\mbs{m}}|\mbs{\theta}), \nonumber
        \end{align}
        where we have dropped the $(w), (m)$ superscript as the model types are
        evident from their inputs. We applied
        \autoref{eq:mixture_weight_normalization} to require only a single
        mixture coefficient.  Note that the stream and background PDFs can
        consist of further components. This will become important for our
        characterization of the stream \stream{GD-1}, which appears spatially
        bifurcated. We note this as an implementation detail since all
        stream-related components (cold, extra-tidal, perturbed) and all other
        components (stellar halo, globular clusters, etc) may be grouped into
        multi-component \autoref{eq:stream_and_bkg_prob} \textit{stream} and
        \textit{background} models.  The stream and background models are
        included graphically in \autoref{fig:PGM}, where the left plot shows the
        stream model, and the right the background.

        The probability of a being a stream member is then simply
        \begin{equation}\label{eq:membership_prob}
            \pdf\!\left(S | \nth{\mbs{w}}, \nth{\mbs{m}}, \mbs{\theta} \right)
            = \frac{f \Spdf(\nth{\mbs{w}}|\mbs{\theta}) \Spdf(\nth{\mbs{m}}|\mbs{\theta}) }{ \pdf(\nth{\mbs{w}}, \nth{\mbs{m}} | \mbs{\theta})}.
        \end{equation}

        Hereafter, when we refer to ``membership probability,"
        \autoref{eq:membership_prob} provides the formal definition. We nuance
        this definition by considering missing feature dimensions in the
        following subsection.

    \subsection{Missing Phase-Space Observations}
    \label{sub:method:missing_data}

        Streams have a diversity of measurement coverage: several have full 6-D
        astrometrics as well as photometry and chemical abundances available
        \citep[e.g.,][]{Koposov+2019, Antoja+2020, Li+2022}, while most have
        partial coverage, lacking some feature dimensions.  Most streams
        detected by \Gaia \citep{Gaia2016, Gaia2023}, for instance, have proper
        motions but not radial velocities nor high signal-to-noise parallax
        measurements.  Follow-up surveys, like \dataarchive{S5} \citep{Li+2019},
        add many feature dimensions, however only for a select set of streams.
        Given the diversity of coverage it is important that the likelihood
        model -- \autoref{eq:membership_prob} -- account for missing data.

        The method of treatment of missing data depends on the cause for that
        data to be missing.  Broadly, there are two categories of causes:
        randomness and systematics.  Consider a dataset with all features
        measured but with a random subset of features masked.  One possible
        approach is to attempt to impute the missing features, then feed the
        full-featured data to the likelihood model. Alternatively, the
        likelihood model might use only the present features, ignoring those
        missing. If however, features are missing for systematic reasons, due to
        a selection function, then this approach will introduce systematic bias.
        The way to avoid bias where the data is incomplete due to the selection
        function is to know that selection function. Otherwise the probabilities
        will not be properly normalized.  Selection functions are often very
        challenging to characterize, so it is routine and convenient to simply
        cut the data where the effects of the selection function become
        significant.  Not including such affected data moves the contribution of
        the selection function from a prior to the evidence.  Our models compute
        likelihoods under a specific dataset, and \autoref{eq:membership_prob}
        is a ratio of the probabilities, so the removed selection functions are
        not expected to be impactful.

        We adopt the approach of modeling the data masks as randomly-distributed
        and not systematic. We do not impute missing data, instead the per-star
        likelihood is modeled with a distribution with the same dimensionality
        as the data vector.  For example, a star with a 6D astrometric
        measurement will be modeled with a full 6D distribution, while a star
        missing its radial velocity measurement will be modeled with a
        dimensionally-reduced (5D) form of the 6D distribution. Using
        dimensionally reduced distributions ensure that missing phase-space
        dimensions do not amplify nor suppress the likelihood.  Notwithstanding,
        we include a quality flag on our calculations indicating the number of
        missing features.  We leave incorporating priors on the data mask
        distribution to future work. 


    \subsection{Pre-training distributions} 
    \label{sub:method:pre-training_distributions}

        In \autoref{sub:method:likelihood_setup} we set up the general
        mathematical framework of MDNs, then specified the models on two axes:
        the component and the data type. Broadly, the components are stream
        versus background and the data types are astrometric and photometric.
        For some components, e.g. the stream, the data density will be well fit
        by an analytic function.  For others, no analytic density provides a
        realistic description of the data. In this work we take a data-driven
        approach, and represent non-analytic distributions using a normalizing
        flow model.

        Normalizing flows provide a flexible description of a probability
        density, by transforming samples generated from a simple base
        distribution (i.e., a Gaussian) to the more complicated target
        distribution \citep{EstebanVanden2010, Rippel2013, RezendeMohamed2015}.
        The transformations are typically non-linear outputs of a neural
        network, whose parameters are optimized until the target density
        accurately characterizes the data. The neural network transformation is
        differentiable, so that jacobian factors can be combined to ensure that
        the target distribution remains a valid probability density (i.e.,
        integrates to one) \citep{Kobyzev+2019}.  For the neural networks we
        regularize the data (minus the mean, divided by the standard deviation
        per feature column); thus we correct the overall normalization of the
        flow by the product of the jacobian of the regularization operation.

        Our implementation is as follows. For a given stream, we restrict the
        data to some field characterized by a polygon drawn around the stream.
        Because our density modeling is aimed at stream characterization rather
        than discovery, we assume that the stream is sufficiently well known so
        that it can be masked out as a simple cut in (e.g.) $\phi_1-\phi_2$
        coordinates. With the stream masked out, we then fit a normalizing flow
        to the data. Because the distribution can evolve with $\phi_1$, we fit a
        \emph{conditional} normalizing flow, $q_{\rm flow}(\mathbf{x}|\phi_1)$.
        Our central assumption with this model is that the distribution for
        background stars in the masked region are quantitatively similar to
        those in the unmasked region. For a sufficiently narrow mask we expect
        this to be the case, and find that this assumption does not introduce
        bias for the streams considered in this work.

        Once the normalizing flow, $q_{\rm flow}\left(\mathbf{x} |
        \phi_1\right)$, is fit to the data, its parameters are fixed while the
        other components of the model begin training. Therefore, the normalizing
        flow can be rapidly trained and computed as a pre-processing step.

        It is easy to evaluate the log-likelihood of a normalizing flow at a
        point $\mbs{x}$ when all feature dimensions are present. When one (or
        more) feature dimension is missing, however, evaluating the marginal
        log-likelihood involves integrals over the flow. While the procedure is
        straightforward it is computationally expensive. As an alternative, we
        conservatively set the log-likelihood to $0$, the maximum value,
        overemphasizing the flow's contribution to the background model.
        Conceptually, up-weighting the background model decreases the
        false-positive rate in stream membership identification, at the cost of
        increasing the false-negative rate.  Another approach, which we leave to
        a future work, is to construct normalizing flows of each marginal
        distribution, e.g. $q_{\rm flow}\left(\{x_i\}_{i \in \mbs{I}_F, i \neq
        {j, k}} | x_j, x_k, \phi_1\right)$, and select for each $\nth{\mbs{x}}$
        the correct flow given the present features.


    \vspace{10pt}
    \subsection{Determination of the Mixture Weight} \label{sub:method:mixture_weight}

        The mixture weight $f_q(\phi_1)$ as used in \autoref{eq:general_model}
        is characterized as the fraction of stars belonging to the mixture
        component $q$ over an increment $[\phi_1, \phi_1 + \dif{\phi_1}]$. Thus,
        the mixture weight encodes information about the linear density of the
        stream, which has been shown to depend on substructure in the galaxy
        \citep{Siegal-GaskinsValluri2009, Yoon+2011}. The dependence of the
        mixture weight on the linear density is captured by the expression
        \begin{equation}\label{eq:fraction_param}
            f_q\left(\phi_1\right) = \frac{n_q\left(\phi_1\right)}{n_{\rm tot}\left(\phi_1\right)},
        \end{equation}
        where $n_q\left(\phi_1\right)$ is the number density of stars in
        component $q$ (i.e., the number of stars belonging to component $q$ per
        $\phi_1$ interval), and $n_{\rm tot}$ is the total number density of
        stars in the same interval (including those in component $q$). The
        linear  density for component $q$ is then entirely specified by the
        function $n_q(\phi_1)$.

        The mixture model specified in \autoref{eq:general_model} only models
        the weight parameter $f_q$, and not the components of the ratio in
        \autoref{eq:fraction_param}. However, the linear density $n_q(\phi_1)$
        can be readily obtained from $f_q$, provided that $n_{\rm
        tot}\left(\phi_1\right)$ can be calculated. Importantly, this function
        does not depend on the parameters of the on-stream or off-stream model.
        Instead, $n_{\rm tot}(\phi_1)$ is simply proportional to the number of
        stars in a small $\phi_1$ increment. We therefore model $n_{\rm
        tot}\left(\phi_1\right)$ as a post-processing step using a single
        normalizing flow in $\phi_1$. 

        In particular, for a given field the distribution of $\phi_1$
        coordinates is specified by the set $\{\phi_1\}_f$. A normalizing flow
        is trained to this dataset, and will satisfy $\int d\phi_1
        P_{\phi_f}(\phi_1) = 1$ by construction.  We can then obtain the linear
        stream density with
        \begin{equation}
            n_q\left(\phi_1\right) = f_q\!\left(\phi_1\right) \, n_{\rm tot}\!\left(\phi_1\right)  \equiv f_q\!\left(\phi_1\right) P_{\phi_1}\!\!\left(\phi_1\right) .
        \end{equation}


    \subsection{Astrometric Model} \label{sub:method:astrometric_model}

        \subsubsection{On-Stream} \label{ssub:method:astrometric_model:on_stream}
    
            The N-dimensional astrometric track is modeled with a single neural
            network, parameterized as a function of $\phi_1$.  The neural
            network output for a single stream component is the mean location of
            the stream track in each dimension, $\astroM{\mbs{\mu}}(\phi_1)$,
            and its associated covariance matrix
            $\astroM{\mbs{\Sigma}}(\phi_1)$. The probability for the data is
            then
            \begin{equation}
                \cmp{S,w}{\pdf}(\nth{\mbs{w}} | \mbs{\theta})
                \!= \mcal{N}_F \big(\nth{\mbs{w}} | \astroM{\mbs{\mu}}\!(\phi_1), {\cmp{w}{\nth{\mbs{\Sigma}}}}^2 \! + \!  \astroM{\mbs{\Sigma}}^2\!\left(\phi_1\right) \!\big),
            \end{equation}
            where $\astroM{\mbs{\theta}}$ is the set of neural network
            parameters that define $\astroM{\mbs{\mu}}$ and
            $\astroM{\mbs{\Sigma}}$; and $\cmp{w}{\nth{\mbs{\Sigma}}}$ the data
            uncertainties.  In the simplest case, we take
            $\cmp{w}{\nth{\mbs{\Sigma}}}, \astroM{\mbs{\Sigma}}$ to be diagonal,
            consisting of both intrinsic dispersion for each dimension of
            $\mbs{w}$ and the data uncertainties added in quadrature. The data
            uncertainties $\cmp{w}{\nth{\mbs{\Sigma}}}$ are elements of
            $\mbs{\delta}\mbf{W}$, the dataset of astrometric uncertainties.

            In practice, we work with truncated Gaussians, where the truncation
            is set by the size of the field in each astrometric dimension.
            Details of the truncated Gaussian are included in
            \autoref{app:sub:normal_distribution}. Truncating to the field
            ensures that the probability density is zero wherever there is no
            data. This has an added benefit of correctly normalizing the
            Gaussian distribution, which is not compact (has non-zero
            probability over all space) and would otherwise not integrate to one
            within the field.

            As a MDN, we implement the Gaussian distribution parameters with a
            multi-layered feed-forward neural network. The number of layers
            required depends on the maximum dimensionality of the Gaussian, and
            whether the neural network employs regularization techniques like
            dropout. We use alternating layers of linear and tanh units. Details
            of the network architecture are discussed in
            \autoref{sub:methods:model_implementation}.

            If a particular observation has missing phase-space coordinate(s) $f_k$, we reduce
            the dimensionality of the Gaussian to that of the data-vector.
            Namely, for $\mbs{I}_{\tilde{F}} \equiv \mbs{I}_{F} \backslash
            \{f_k\}$
            \begin{align}
                \nth{\tilde{\mbf{w}}} &\mapsto \nth{\tilde{\mbf{w}}} = \{ w_{n,f} | f\in \mbs{I}_{\tilde{F}}\}, \nonumber\\
                \astroM{\mbs{\mu}} &\mapsto \astroM{\tilde{\mbs{\mu}}} = \{ \mu_{f} | f\in \mbs{I}_{\tilde{F}}\}, \nonumber\\
                \astroM{\mbs{\Sigma}} &\mapsto \astroM{\tilde{\mbs{\Sigma}}} = \{ \Sigma_{i,j} | i,j \in \mbs{I}_{\tilde{F}} \}, & \nonumber
                \intertext{such that,}
                \Spdf(\nth{\mbs{w}} & | \astroM{\mbs{\theta}}(\phi_1))
                    \mapsto \Spdf(\nth{\tilde{\mbs{w}}} | \astroM{\tilde{\mbs{\theta}}} (\phi_1))\,.
            \end{align}


        \subsubsection{Off-stream}

            \label{ssub:method:astrometric_model:off_stream} We now outline the
            off-stream astrometric model. We take a similar approach to the
            on-stream astrometric model discussed in
            \autoref{ssub:method:astrometric_model:on_stream}, though we
            typically do not characterize the backgrounds with a Gaussian
            distribution. Instead, we utilize a range of distributions (i.e.,
            exponential, skew-normal) discussed in \autoref{app:distributions}.
            The parameters of the user-specified background distributions are
            themselves neural network outputs, such that any $\phi_1$ ``slice"
            of the background should be characterized by some analytic density,
            but the evolution of the backgrounds with $\phi_1$ can be complex.
            Details of the network architecture are discussed in
            \autoref{sub:methods:model_implementation}.
    
            Let $\Bcmp{\fth{\mbs{\theta}}}(\phi_1)$ represent the neural network
            which outputs a vector of parameters characterizing the background
            distribution at a given $\phi_1$ for astrometric dimension $f$. The
            number of outputs for the background model is equal to
            $\mathrm{dim}\left(\fth{\Bcmp{\mbs{\theta}}}\right)$ (where $f=0$
            corresponds to $\phi_1$, which we do not model). At a given
            $\phi_1$, the probability density for the astrometric dimension
            $f>0$ is $\Bcmp{\fth{\pdf}}\left(\mbs{w}_{n,f} |
            \Bcmp{\fth{\mbs{\theta}}}(\phi_1) \right)$. We treat each background
            component dimension as independent, allowing us to write the
            likelihood as a product over the astrometric dimensions:
            \begin{equation} \label{eq:astrometric_model_off_stream_probability}
                \pdf^{(B)}\left(\nth{\mbs{w}} | \mbs{\theta}(\phi_1) \right) = \prod_{f\in \mbs{I}_{\tilde{F}}} \pdf_f^{(B)}\left(\mbs{w}_{n,f} | \Bcmp{\fth{\mbs{\theta}}}(\phi_1) \right)\,,
            \end{equation}
            where $\mbs{I}_{\tilde{F}}$ is the set of indices of the features,
            omitting missing phase space dimensions.
        


    \subsection{Photometric Model} \label{sub:method:photometric_model}

        \subsubsection{On-Stream} \label{ssub:method:photometric_model:on_stream}

            In photometric coordinates we model the stream as a
            single-population isochrone with the possibility of a non-zero
            distance gradient. We accomplish this by modeling the distance of
            the isochrone as a neural network output, parameterized as a
            function of $\phi_1$. The result is a distance track along the
            stream, estimated from the variation in its isochrone with $\phi_1$. 
            
            At any given $\phi_1$, the isochrone is modeled in absolute
            magnitudes and parameterized by the normalized stellar mass, labeled
            $\gamma \in (0, 1]$, in the modeled mass range. For example,
            increasing $\gamma$ corresponds to movement up the main sequence
            towards the red giant branch. We denote the isochrone as
            $\mcal{I(\gamma)}$, and \package{StreamMapper} permits any such
            distribution. The intrinsic dispersion around the isochrone is
            $\Sigma_\mcal{I}(\gamma)$. Since the isochrone is in absolute
            magnitudes, but the data are in apparent magnitudes, it is necessary
            to shift the isochrone by a distance modulus to the predicted
            distance of the stream, labeled  $\mu(\phi_1)$, with intrinsic
            stream distance variation $\sigma_\mu(\gamma)$.
            
            For the $n$-th star, the photometric model is:
            \begin{multline} \label{eq:photometric_probability_point}
                \cmp{S,m}{\pdf}(\mbs{m_i}, \gamma | \mbs{\theta}(\phi_1)) 
                \\ \sim \mcal{N}(\nth{\mbs{m}} | \{\mcal{I(\gamma)} + \mu(\phi_1), \mbs{\Sigma}_{\mcal{I}}^2(\gamma) + \mathds{1} \sigma_\mu^2(\phi_1) + {\cmp{m}{\nth{\mbs{\Sigma}}}}^2 \}) \\ \times \prior(\gamma, \nth{\mbs{m}}, \phi_1),
            \end{multline}
    
            where $\cmp{m}{\nth{\mbs{\Sigma}}}$ is the data covariance and
            $\prior(\gamma, \nth{\mbs{m}}, \phi_1)$ encodes both the present-day
            mass function of the stream, as well as any observational
            constraints (e.g., $g < 20$ mag).

            For a given star, we are interested in its probability of belonging
            to the stream's isochrone. Therefore, we marginalize over $\gamma$
            to find:
            \begin{small}
            \begin{align} \label{eq:photometric_probability}
                &\cmp{S,m}{\pdf}(\nth{\mbs{m}} | \mbs{\theta}(\phi_1)) = \\
                &\ \int\limits_{\gamma=0}^{1} \mcal{N}(\nth{\mbs{m}} | \{\mcal{I(\gamma)} + \mu(\phi_1), \mbs{\Sigma}_{\mcal{I}}^2(\gamma) + \mathds{1}\sigma_\mu^2(\phi_1) + {\cmp{m}{\nth{\mbs{\Sigma}}}}^2 \}) \nonumber
                \\
                &\qquad \times \prior(\gamma, \nth{\mbs{m}} | \phi_1) \rm{d}\gamma. \nonumber
            \end{align}
            \end{small}
            Then, for a given $\phi_1$ we can obtain a distribution over the
            distance modulus using 
            \begin{align}
                \mu_{\rm sample}\left(\phi_1\right) \sim \mathcal{N}\left(\mu(\phi_1), \sigma^2_\mu(\phi_1) \right). 
            \end{align}
            For each sample, we can convert the distance-modulus distribution
            (and error) to the distance track using 
            \begin{align}
                \rm{dist}\left(\phi_1\right) &= 10^{\frac{1}{5}\mu_{\rm sample}(\phi_1)+1} & \quad\unit{pc}, \\
                \delta \rm{dist}(\phi_1) &\simeq \frac{\ln{10}}{5} \rm{dist}(\phi_1) \ \sigma_\mu(\phi_1) & \quad\unit{pc}.
            \end{align}

            Let $\cmp{S,m}{\mbs{\theta}}(\phi_1)$ represent the feed-forward
            neural network which outputs a vector of parameters characterizing
            the stream's photometric distribution at a given $\phi_1$. The
            network has 2 outputs -- $\mu, \sigma_\mu$ -- as we only consider
            the distance track of the isochrone, holding fixed stellar
            population parameters internal to modeling the isochrone, such as
            its age and metallicity.  In principle we might include parameters
            of the isochrone in the model, now $\mcal{I}(\mbs{\theta}, \gamma)$,
            where $\theta = {a, Z, ...}$ and any other generating parameters of
            the isochrone.  However, at a detailed level stellar streams are
            generally poorly fit by theoretical isochrones, at least ones with
            realistic ages and metallicities.  Consequently including the age
            and metallicity as model parameters does not meaningfully contribute
            to constraining results on the stream's stellar population.
            Moreover, age and metallicity are partially degenerate with the
            distance modulus, working counter to the goal of fitting a distance
            track to the stream. An alternate approach is to use data-driven
            isochrones instead of theoretic ones.   \package{StreamMapper}
            allows any isochrone from any theoretic or data-driven library to be
            used, so long as the isochrone may be described as a curve
            $\mcal{I}(\gamma, \phi_1)$ with dispersion
            $\Sigma_{\mcal{I}}(\gamma, \phi_1)$. In this work we use MIST
            \citep{Dotter2016, Choi+2016} isochrones for their convenience, but
            recognize their limitations as compared to data-driven isochrones.
            We defer using data-driven isochrones and constraining ages and
            metallicities to a future work.

            The photometric probability, \autoref{eq:photometric_probability},
            encodes the photometric distribution -- the ``track" -- of the
            isochrone model, as well as the relative probability of points along
            the track. The latter is the term $\prior(\gamma, \nth{\mbs{m}} |
            \phi_1)$, which describes all priors on both the isochrone as well
            as survey constraints and selection effects. We assume that the
            properties of the isochrone are conditionally independent from
            observational constraints; namely
            \begin{equation} \label{eq:isochrone_prior_components}
                \prior(\gamma, \nth{\mbs{m}} | \phi_1) = \prior_{\cal{I}}(\gamma | \phi_1) \prior_{obs}(\nth{\mbs{m}} | \phi_1)
            \end{equation}
            where $\prior_{\mcal{I}}$ and $\prior_{obs}$ are the isochrone and
            observational priors, respectively.
            
            The second term in \autoref{eq:isochrone_prior_components} --
            $\prior_{obs}(\nth{\mbs{m}} | \phi_1)$ -- encodes all observational
            effects in the photometric dataset. Some effects are easily modeled,
            e.g. instruments have well-characterized magnitude limits, both
            faint and bright. Other effects are not easily modeled, e.g, complex
            location-dependent completeness issues \citep{GaiaCompleteness}.
            Accounting for the completeness is very challenging. In practice we
            avoid the contribution of completeness to $\prior_{obs}$, masking
            data for which it is relevant. \Gaia, for instance, is complete from
            $12 < G < 17$ and has high completeness up to $G \approx 20$, thus
            we mask $G > 20$.  The masks are easily modeled as magnitude limits
            and replace the complex completeness model in $\prior_{obs}$.
            \autoref{sub:method:missing_data} explains how masking is
            implemented in the models, though in this case every feature is
            masked, not only a few feature dimensions. With every featured
            masked, the photometric model in
            \autoref{eq:photometric_probability} reduces from an
            $F_m$-dimensional Gaussian to a 0-D Gaussian, which is a null
            distribution and does not enter into the likelihood.  In photometric
            coordinates completeness depends on a stars apparent magnitude;
            viewed in astrometric coordinates, the completeness cannot be
            described so simply. \Gaia, for instance has scanning law residuals
            in the density field. When photometric coordinates are masked due to
            completeness, we do not mask the corresponding  astrometric
            features, so that the stars membership probability is determined
            solely by the astrometric model. This practice allows us to include
            stars with good astrometric measurements, not penalizing those stars
            for which the photometric model is insufficient.  We examine the
            results to confirm that the scan pattern does not impact the
            membership distribution.

            The first term in \autoref{eq:isochrone_prior_components} --
            $\prior_{\mcal{I}}$ -- encodes the point-wise amplitude of the
            isochrone.  Physically, this amplitude is the the present-day mass
            function (PDMF). The PDMF $\prior_{\cal{I}}(\gamma, \phi_1)$ is a
            joint distribution, depending on $\gamma$ since it is the
            $\gamma$-dependent distribution amplitude, and also on $\phi_1$.
            This latter dependence is confusingly termed the ``mass function of
            the stream", which is the expected variation in the stellar mass
            distrubion along the stream, arising due to e.g. mass segregation in
            the progenitor \citep{WebbBovy2022}. To date this variation along
            the stream has not been observed, so in practice we drop the
            $\phi_1$ dependence -- $\prior_{\cal{I}}(\gamma, \phi_1) =
            \prior_{\cal{I}}(\gamma)$ -- assuming that the PDMF is constant
            along the stream.

            In the
            \href{https://github.com/GalOrrery/stream_mapper-pytorch}{\package{StreamMapper}}
            we offer a few ansatz for the PDMF, though any user-defined function
            may be used instead. The default assumption, known to be
            non-physical, is a uniform distribution over $\gamma$.  Similarly
            non-physical are truncated uniform distributions, considering only a
            segment of the isochrone, e.g. the main sequence turnoff (MSTO). We
            include also a Kroupa initial mass function (IMF)
            \citep{Kroupa2001}. The Kroupa IMF, and all physically-motivated
            IMFs, have a much larger fraction of low-mass stars than high-mass
            stars, the latter being less likely to form.  This is observed in
            streams, particularly their progenitors. However the Kroupa IMF, and
            most out-of-the-box IMFs, are poor fits to the observed PDMF as
            stream progenitors like globular clusters are extremely old
            ($\sim10$ Gyr) and may be dynamically evolved
            \citep{GrillmairSmith2001} while an IMF is the zero-age
            distribution.  How then should the PDMF $\prior_{\cal{I}}$ be
            treated?  One approach to the PDMF is to use a more informed
            distribution, for example time-evolving an IMF, either analytically
            or by simulation, to the age of the stream's stellar population.
            However, for many streams, e.g. those lacking known progenitors, the
            PDMF is observationally poorly constrained.  In short, using any
            fixed distribution is simple in implementation, however the choice
            of distribution is not.

            Rather than using a fixed \textit{a priori} ansatz, another approach
            to the PDMF is to incorporate it into the model, parameterized by
            some distribution and fit simultaneously to the distance modulus and
            other $\cmp{m}{\mbs{\theta}}$.  We find this, like incorporating $a,
            Z$ into the isochrone model, to be interesting avenues by which to
            develop the model. However the resulting significant increase in
            model flexibility necessitates careful regulation, and we leave this
            to a future work.

            The last approach to the issue of the PDMF, and the one we take in
            this work, is to render it unimportant.  With fixed age,
            metallicity, and other isochrone model parameters, the primary
            purpose of the isochrone is to allow the model to determine the
            distance modulus of the stream. For this, only a portion of the
            stream need be modeled.  For streams with observable main sequence
            turn-offs, like \stream{GD-1} in \Gaia, this portion of the
            isochrone has both many stars and a relatively small range in
            masses. Thus, we restrict the mass range of the isochrone such that
            we are modeling only a region of interest over which the PDMF does
            not meaningfully contribute to the final result.


        \subsubsection{Linking to Astrometrics: the Distance Modulus and the Parallax} \label{ssub:method:linking_to_astrometrics}

            By modeling jointly the astrometrics and photometrics we may tie
            together components of the models.  In particular, the distance
            modulus in the photometric model and the parallax in the astrometric
            model are both measures of the distance track of the stream. We
            model in the same coordinates as the data, and thus prefer parallax
            and distance modulus over the distance for each model type.  We
            introduce a prior to connect these components of the models:
            \begin{align}
                \prior(\{\mu_I, \mu_\parallax \} | \phi_1) &= \delta( \rm{dist}(\mu_I) - \rm{dist}(\parallax) )(\phi_1), \label{eq:distance_track_prior}
            \end{align}
            where $\delta$ is the Dirac delta function \citep{Dirac1947}, and
            enforces that the distance modulus track and parallax track must be
            equal (when converted to the actual distance) as a function of
            $\phi_1$.  The widths are similarly constrained. The distance and
            first-order error conversion are given by
            \begin{align}
                \mu_I(\parallax) &= 10 - 5\log_{10}(\parallax) \\
                \sigma_I(\sigma_\parallax)  &\approxeq \frac{5}{\ln{10}} \mu_I \lvert\sigma_\parallax\rvert.
            \end{align}
    
            In practice, at the distance of many streams the errors in the
            parallaxes are of order 1 and we exclude the feature from the model.

    
        \subsubsection{Off-stream}
        \label{sub:method:photometric_model:off_stream}

            The distribution of background stars in color (i.e., $g-r$) and
            magnitude (i.e., $g$) is a complicated function of location in the
            galaxy. Even within a field centered on the stream of interest, the
            color-magnitude diagram (CMD) is a combination of stellar
            populations distributed over a range in distances. Because of this,
            we do not find an analytic density that describes the data and use
            instead a normalizing flow, discussed in
            \autoref{sub:method:pre-training_distributions}.  Because the
            background CMD can evolve with $\phi_1$, we fit a \emph{conditional}
            normalizing flow, $q_{\rm flow}(\mathbf{m}|\phi_1)$, so that the
            color-magnitude diagram is conditionally dependent on position along
            the stream. In practice, we model magnitudes and not colors (i.e.,
            $\mathbf{m} = (g,r)$ rather than $g \ \mathrm{and} \ g-r$.
            Otherwise, the two arguments will always be covariate). The
            normalizing flow approach makes modeling the photometry relatively
            straightforward.

            In photometry the intrinsic width of stellar streams tends to be
            quite small, with correspondingly small $\sigma_\mu$ (see
            \autoref{eq:photometric_probability}). Therefore the effective
            region of the isochrone model is well determined and we do not find
            the marginal log-likelihood approach discussed in
            \autoref{sub:method:pre-training_distributions} to be impactful.



    \subsection{Priors} \label{sub:methods:priors}

        In \autoref{sub:method:likelihood_setup} we introduced the likelihood
        setup. In \autoref{ssub:method:photometric_model:on_stream} we added
        priors in the form of the isochrone's PDMF and photometric observational
        constraints. In this section we introduce and develop more priors that
        are important for training our Bayesian Mixture Density Network.

        \subsubsection{Parameter Bounds} \label{ssub:methods:priors:parameter_bounds}

            For each parameter in $\mbs{\theta}$ of the model we implement
            priors on its range.  Uninformative, e.g. uniform $\mcal{U}(a, b)$,
            priors are often assumed where little is known about a parameter
            (except that $(a,b]$ contain the correct $\theta$ and its
            distribution). Where more is known about $\theta$, for instance that
            it is a categorical, more informative priors like the Dirichlet
            distribution \citep{Christopher2016} may be used. Our code,
            \package{StreamMapper}, permits any user-defined prior, and we
            include $\mcal{U}(a, b)$.  The default choice of prior and bounds
            $\mcal{U}$ set to the limits of the dataset in each feature
            dimension.

            We implement a hard-coded uniform prior on a neural network output
            by applying a scaled logistic function of the form
            \begin{equation}\label{eq:scaled_logistic}
                  \sigma(x, \{a,b\}) = a + (b - a) * \sigma(x)
            \end{equation}
            Therefore, we pair every parameter bound with a matching bound in
            the network architecture. Further details of this practice are
            explained \autoref{app:nns_and_priors}.

            The parameter bounds may be independent of $\phi_1$, as is implicit
            to the above discussion, or conditionally dependent and vary as any
            user-defined function of $\phi_1$.  Alternatively, prior parameters
            may be separated into pieces, dealing with different modeling needs,
            e.g. defining broad parameter bounds as was done in this section.
            We introduce a different piece subsequently.


        \subsubsection{Priors on the Stream Track} \label{sub:methods:priors:track_region_prior}

            We now discuss how our model incorporates known information about
            the track of the stream.  Our density modeling is aimed at
            characterization of \textit{known} streams, not stream discovery.
            From previous studies, the broad track of a stream is known for
            significant portions of its length \citep[e.g. see the atlas
            in][]{Mateu2022}. It makes sense, then, to provide this information
            to the model. Care must be taken that this prior information does
            not dominate the model.  We introduce a ``guide'' prior to encourage
            the stream model to pass through a user-defined guide region, while
            also remaining compatible with the data.  We use a ``split''-Normal
            prior: a Gaussian split and separated at the peak and pieced back
            together with a flat region connecting the halves.  The
            un-normalized PDF is given by:
            \begin{equation}
                \pdf(x,\mu,\sigma,w) \propto \begin{cases} 
                   \Exp{-\frac{1}{2}\left(\frac{x-(\mu-w)}{\sigma}\right)^2} & \phantom{\mu - w <}\ x < \mu - w \\
                    1 & \mu - w \leq x < \mu + w \\
                    \Exp{-\frac{1}{2}\left(\frac{x-(\mu+w)}{\sigma}\right)^2} & \mu + w < x
                \end{cases}.
            \end{equation}
            Since the derivative at the peak of a Gaussian is $0$, the piecewise
            PDF is smooth up to the first derivative and can be used with
            gradient-based optimizers.

            The Gaussian portions of the prior encourage the network towards the
            central region $\mu - w < x < \mu + w$. Because the prior is
            equi-probable in the guided region, the stream may lie anywhere as
            preferred by the data and the stream model.  It is convenient to
            parameterize, with $\tau = \frac{1}{\sigma}$. Namely,
            \begin{small}
            \begin{equation}
                \ln \pdf(x,\mu, \tau, w) \propto \tau
                \begin{cases} 
                    -\left(x-(\mu-w)\right)^2 & \phantom{\mu - w <}\ x < \mu - w \\
                    0 & \mu - w \leq x < \mu + w \\
                    -\left(x-(\mu+w)\right)^2 & \mu + w \leq x
                \end{cases},
            \end{equation}
            \end{small}
            where the additive normalization is
            \\
            $-\ln\left(\sqrt{\frac{\pi}{\tau}}
            \left(\frac{1}{2}\text{erf}\left(2 \sqrt{\tau }
            w\right)+1\right)\right)$.  In this form we see that $\tau$ may be
            increased to strengthen the prior and encourage $\theta$ to lie
            within the desired region. It is important that the region be set
            large enough that we may be confident the prior is not driving the
            final result on small scales. As an added measure, it may useful to
            turn off this prior after the stream lies within the desired guides
            and the learning rate of the optimizer is low enough that the model
            does not escape the solution minima.



    \subsection{Model Implementation} \label{sub:methods:model_implementation}

        In previous sections we discussed the mathematical implementation of
        each component of the Mixture Density Network. Here we make brief
        remarks on the code implementation in \package{StreamMapper} and the
        architecture of the neural networks.

        The neural networks are all multi-layer perceptrons (MLP; built with the
        \package{PyTorch} stack \citep{Pytorch2019}. We wrap the network
        \texttt{pytorch.Module} in a custom framework, allowing the Bayesian
        likelihood, prior, evidence, etc. to be bundled with the network's
        forward step. This framework allows for a high degree of modularity.
        Component models, e.g. the stream model, of the mixture are easily added
        to (and moved from) as items in the larger mixture model. Probability
        methods understand how to traverse the mixture model, calling and
        combining the appropriate methods of the constituent models, so it
        unnecessary to write custom loss functions.  Another benefit of the
        modular framework is the ease of building and experimenting with
        different mixture models. Furthermore, each model component may be saved
        and serialized separately from the other components, allowing for
        portability of the components between models.  For example, if a stream
        has a prominent extra-tidal feature, e.g. the spur of \stream{GD-1}
        \citep{Bonaca+2019}, a fiducial model of just the main stream may be
        trained, then the stream (and background) model components may be saved
        and used as the initialization to the stream (and background) of a more
        complete model that includes the spur.  By using a modular framework,
        the architecture of each neural network may be chosen to best-suit its
        modeling needs: e.g. thin networks for simply varying components.  On
        the other hand, large networks can have useful emergent phenomenon, e.g.
        understanding covariate structures, that cannot be (or are unknown how
        to be) replicated with multiple smaller networks. We choose to use the
        smaller and more tailored models, finding their balance of performance
        and reduced training time well-suited for this work.

        The models permit user-defined priors to be passed at model
        instantiation.  Priors specific to a single model are contained on that
        model, while priors connecting two (or more) models are held on the
        mixture model object. As an example of the latter, see
        \autoref{ssub:method:linking_to_astrometrics}, which connects the
        distance component of the astrometric and photometric stream models.

        The specific architecture of the neural network(s) is set and determined
        by the user. So long as the network has the correct number of inputs
        (e.g. one for $\phi_1$) and outputs (one per parameter), any
        feed-forward network may be used.  Our default choice, and for which we
        provide a helper function, is an MLP with a sequence of \texttt{Linear}
        \!\!\textrightarrow \texttt{Tanh} \!\!\textrightarrow \texttt{Dropout}
        (hereafter \texttt{LTD}) blocks.  Dropout layers are used to prevent
        network co-adaptation and over-fitting. Dropout works by randomly
        zeroing out nodes in the network, decreasing the importance of any
        single node, encouraging instead emergent patterns across the network
        \citep{GalGhahramani2015}.  One model for which the \texttt{LTD} network
        cannot be used is the normalizing flow used in the photometric
        background model (\autoref{sub:method:photometric_model:off_stream}).
        Here we use the \package{zuko} package for normalizing flows, with a
        conditional diagonal normal base distribution and composite transform
        network of a reverse permutation and masked affine auto-regressive
        transform layers. As noted in our data availability statement, all the
        codes used in this paper are bundled by
        \href{https://github.com/showyourwork/showyourwork}{\package{showyourwork}}
        \citep{Luger+2021} in a public repository. We will describe for each
        stream model specifics relevant to that stream but refer readers to the
        publicly-available code itself for more implementation details.
        

\section{Results: Mock} \label{sec:results_mock}

    In this section we present an application of our model to a synthetic stream
    with known ground truths, and demonstrate the model's ability to
    characterize stream density variations and membership probabilities in an
    unsupervised manner.

    \subsection{Data Simulation} \label{sub:results_mock:data}

        We generate synthetic stream observations by sampling from simple
        algebraic forms.  In $\phi_1$ we sample from a uniform distribution.
        Non-Poissonian density variations, like gaps, are introduced by adding
        to the uniform distribution two Gaussian distributions centered at
        different values of $\phi_1$ with negative amplitudes and random widths.
        Stream stars are then sampled from this uniform-plus-gaps mixture
        distribution.  In $\phi_2$ we sample from a quadratic in $\phi_1$ with
        normally-distributed $0.15^{\degree}$ scatter.  The distances are
        uniformly sampled from $7$ to $15$ kpc in a linear function in $\phi_1$
        with normally-distributed $0.25 \ \rm{kpc}$ scatter, then transformed to
        the parallax $\parallax$. Even though the stream width is constant in
        the distance the width in parallax is not, which will be used to
        demonstrate that we can recover $\phi_1$ dependent distributions.  The
        number of stream stars is $%
  $1569$\label{output/mock/nstream_variable.txt}\unskip%
$.
        $%
  $13000$\label{output/mock/nbackground_variable.txt}\unskip%
$ background astrometric
        points are generated by sampling from uniform distributions in each
        coordinate.
    
        The synthetic stream's photometry is also simulated using a
  12 Gyr\label{output/mock/isochrone_age_variable.txt}\unskip%
\unskip, [Fe/H] =
  -1.35~dex\label{output/mock/isochrone_feh_variable.txt}\unskip%
 MIST isochrone
        \citep[using][]{brutus}. The isochrone is truncated just short of the
        giant branch, and is sampled from assuming a stream mass function
        similar to \stream{Pal\,5} \citep{GrillmairSmith2001} as well as with an
        intrinsic uniform 0.2~dex scatter orthogonal to the isochrone track.

        The isochrone is shifted by the stream's distance modulus, which is
        derived from the astrometric distance track.  The background photometric
        coordinates are generated as a 2-dimensional Gaussian with non-zero
        covariance. The background is illustrative, if not realistic, in that it
        is a distribution about which we care very little beyond our ability to
        fit with a model.  Thus, the synthetic data scenario described here, and
        presented in \autoref{fig:mock_data_photometric_background_selection},
        generates a mock stream in astrometric coordinates and photometric
        coordinates, with density variations, a distance gradient which is
        consistent in parallax and distance modulus (i.e., magnitudes), all
        superimposed over a $\sim 10$ times larger noisy field of background
        points in each astrometric and photometric dimension. 

    \subsection{Model Specification} \label{sub:results_mock:model}

        \begin{figure}
            \script{mock/plot/photometric_background_selection.py}
            \centering
            \includegraphics[width=1\linewidth]{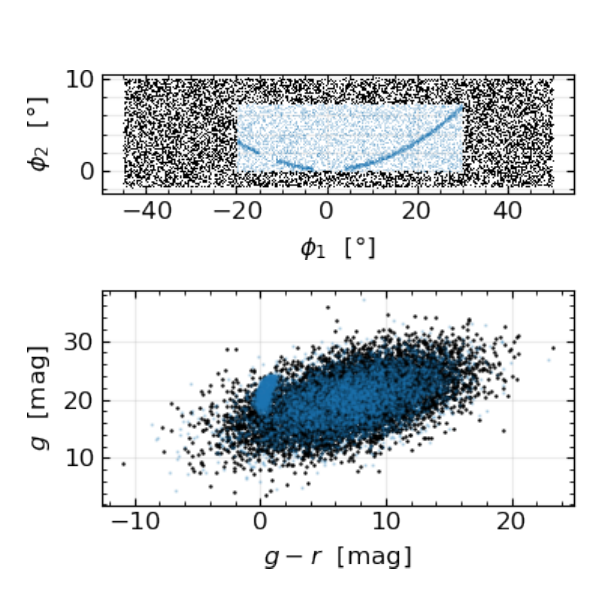}
            \caption{
                \textbf{Top panel:} Astrometric coordinates ($\phi_1$, $\phi_2$)
                for the mock data set.  Light blue points are the stream and
                background within an ``on"-stream region. Black points are in
                the ``off"-stream region used to train the background
                photometric model.
                \textbf{Bottom panel:} Photometric coordinates for the mock data
                set.  Black points are in the ``off"-stream region from the
                region defined in the astrometric coordinates.  Light blue
                points are the stream and background from the on-stream region.
                The stream's isochrone is apparent as an overdensity around
                (g-r, g)$\ \approx (0, 20)$.  The background is an arbitrary
                Gaussian distribution from which the stream must be separated.
            }
            \label{fig:mock_data_photometric_background_selection}
        \end{figure}

        \begin{figure*}[!h]
            \script{mock/plot/results.py}
            \centering
            \includegraphics[width=0.9\linewidth]{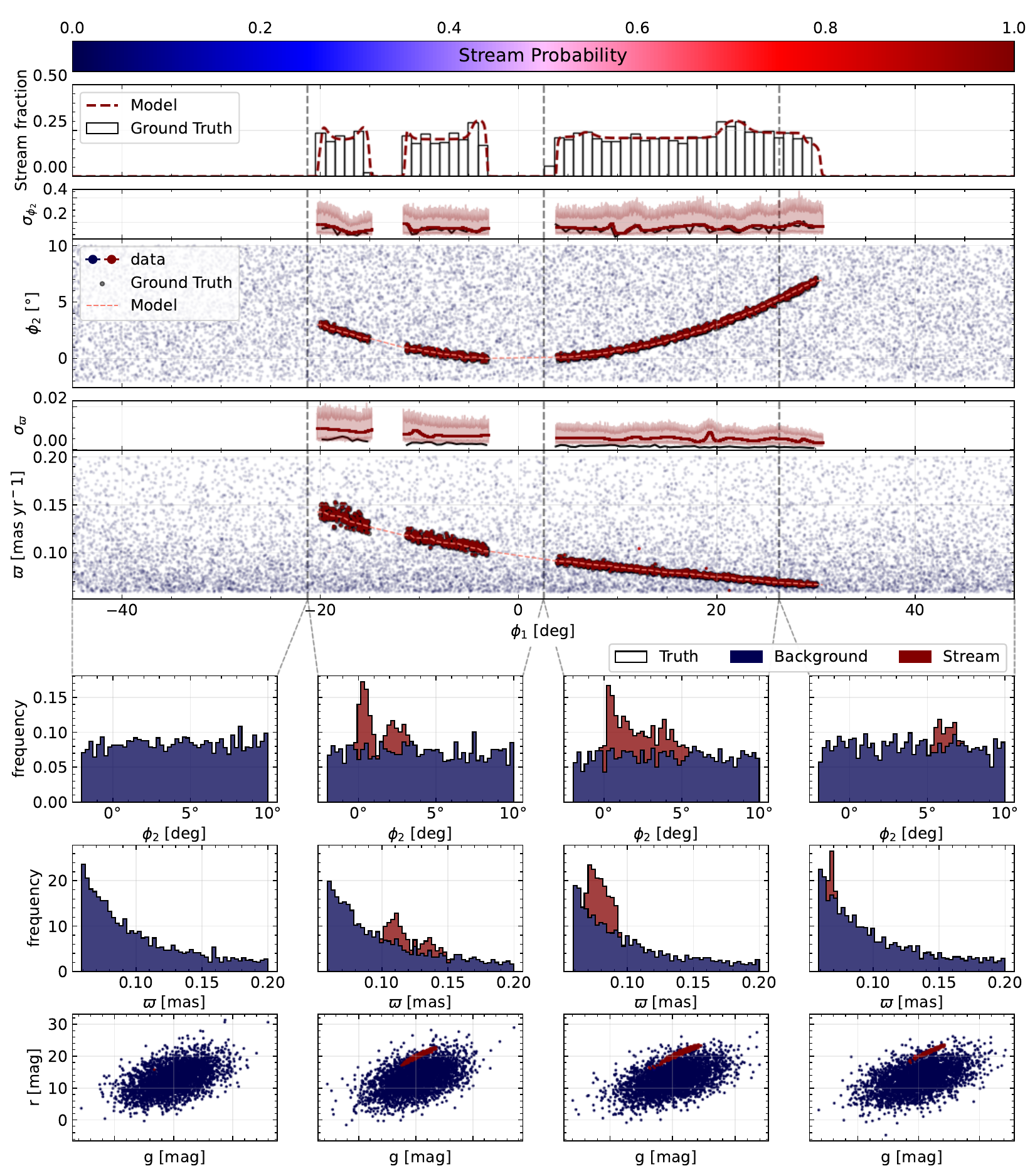}
            \caption{%
                Illustration of the model's performance in recovering a stream
                from synthetic observations. The \textbf{top panel} shows the
                stream mixture coefficient $f(\phi_1)$ (dashed red line)
                predicted by the model, overplotted on the true stream fraction
                (black-outlined histogram).
                In the \textbf{middle two rows} we also show the $\phi_2$ and
                parallax, $\parallax$, tracks as a function of $\phi_1$ and
                their associated widths (labeled as $\sigma_{\phi_2}$ and
                $\sigma_{\parallax}$, respectively). The true values for these
                quantities are shown in gray, while our model prediction and
                uncertainty is show by the dashed red curve and red band, and by
                the color of the data points (color-coded by membership
                probability). The model does an excellent job at identifying the
                ground truth stream stars.
                In the \textbf{bottom three rows} we demonstrate our ability to
                separate the stream from the background both astrometrically and
                photometrically. Histograms are shown for each model component
                (black for the ground truth, blue for the background, red for
                the stream), in 4 $\phi_1$ ranges across the data set.  The
                height of the histogram is the probability-weighted density for
                the collection of stars in each bin. There is nearly perfect
                agreement for the number of stars belonging to each model
                component as a function of $\phi_1$. In the bottom row we show
                photometric plots for each model component in the same 4
                $\phi_1$ ranges across the data set. Points are color-coded by
                stream membership probability. 
            }
            \label{fig:mock_data_result}
        \end{figure*}
    
        We select on-stream and off-stream regions using the known $\phi_1,
        \phi_2$ range of the mock stream.  The upper panel of
        \autoref{fig:mock_data_photometric_background_selection} shows these
        selections in blue and black, respectively.  The lower panel shows the
        on and off-stream selections in photometric coordinates.  In this space
        it is apparent that the on-stream region has two populations, drawn from
        the isochrone and from background points included in the simple $\phi_1,
        \phi_2$ box.  In contrast, the off-stream region has only the the
        background population.  We use the off-stream population to train a
        conditional normalizing flow (see
        \autoref{sub:method:pre-training_distributions}) on the magnitudes $g$
        and $r$ and conditioned on $\phi_1$.  The normalizing flow is now a
        non-parametric distribution characterizing the photometric background,
        including its variation over $\phi_1$.  For this mock stream the
        flexibility of the normalizing flow greatly exceeds necessity: the
        background is drawn from a $\phi_1$-invariant multivariate normal,
        essentially the base distribution of the normalizing flow.  It is
        therefore no surprise that the trained normalizing flow provides an
        excellent characterization of the background.  We will later see that
        the flow performs similarly well on complex distributions.  Having
        trained the background photometric model we proceed to build and train
        the full model.

        Per \autoref{eq:stream_and_bkg_prob} the stream is an independent
        addition of both astrometric and photometric models.  In astrometrics
        the stream is a Gaussian track in $\phi_2$ and parallax $\parallax$,
        using the model defined in
        \autoref{ssub:method:astrometric_model:on_stream}.  We use a 4-layer
        \texttt{LTD} with overall shape ($\phi_1$ input, 4 outputs), where the 4
        outputs are $[\mu_{\phi_2}, \sigma_{\phi_2}, \mu_{\parallax},
        \sigma_{\parallax}](\phi_1)$.  The network is small, but sufficiently
        large to permit 15\% dropout and retain robust predictions.  The
        photometric model is a distance modulus-shifted isochrone, using the
        model outlined in  \autoref{ssub:method:photometric_model:on_stream}. We
        use the same isochrone parameters as were used to simulate the stream,
        including the \citet{GrillmairSmith2001} mass function.  For this mock
        stream, we demonstrate simultaneous astrometric and photometric
        modeling, linking the distance modulus and parallax track, as explained
        in \autoref{ssub:method:linking_to_astrometrics}.

        The total background model has two pieces, an astrometric and a
        photometric model.  As a pre-training phase the photometric normalizing
        flow model was fitted to an off-stream selection and is now held fixed.
        The astrometric model is split in two components, the first is a uniform
        distribution in $\phi_2$, and the second is an exponential distribution
        for $\parallax$.  The uniform distribution does not have any parameters,
        and thus no neural network.  The exponential distribution has only the
        slope parameter and we use a 3-layer \texttt{LTD} with shape (1, 32, 1)
        and 15\% dropout. Background and stream are combined in a mixture model
        whose mixture parameters are a 4-layer (1, 64, 1) \texttt{LTD} network.
        We train the mixture model with an \texttt{AdamW} optimizer with
        learning rate of $5\times10^{-3}$ and a scheduler to periodically and
        temporarily increase the learning rate, encouraging the optimizer
        towards better minima.

    \newpage
    \subsection{Trained Density Model} \label{sub:results_mock:results}

    \begin{figure*}[ht]
        \script{gd1/plot/data_selection.py}
        \centering
        \includegraphics[width=1\linewidth]{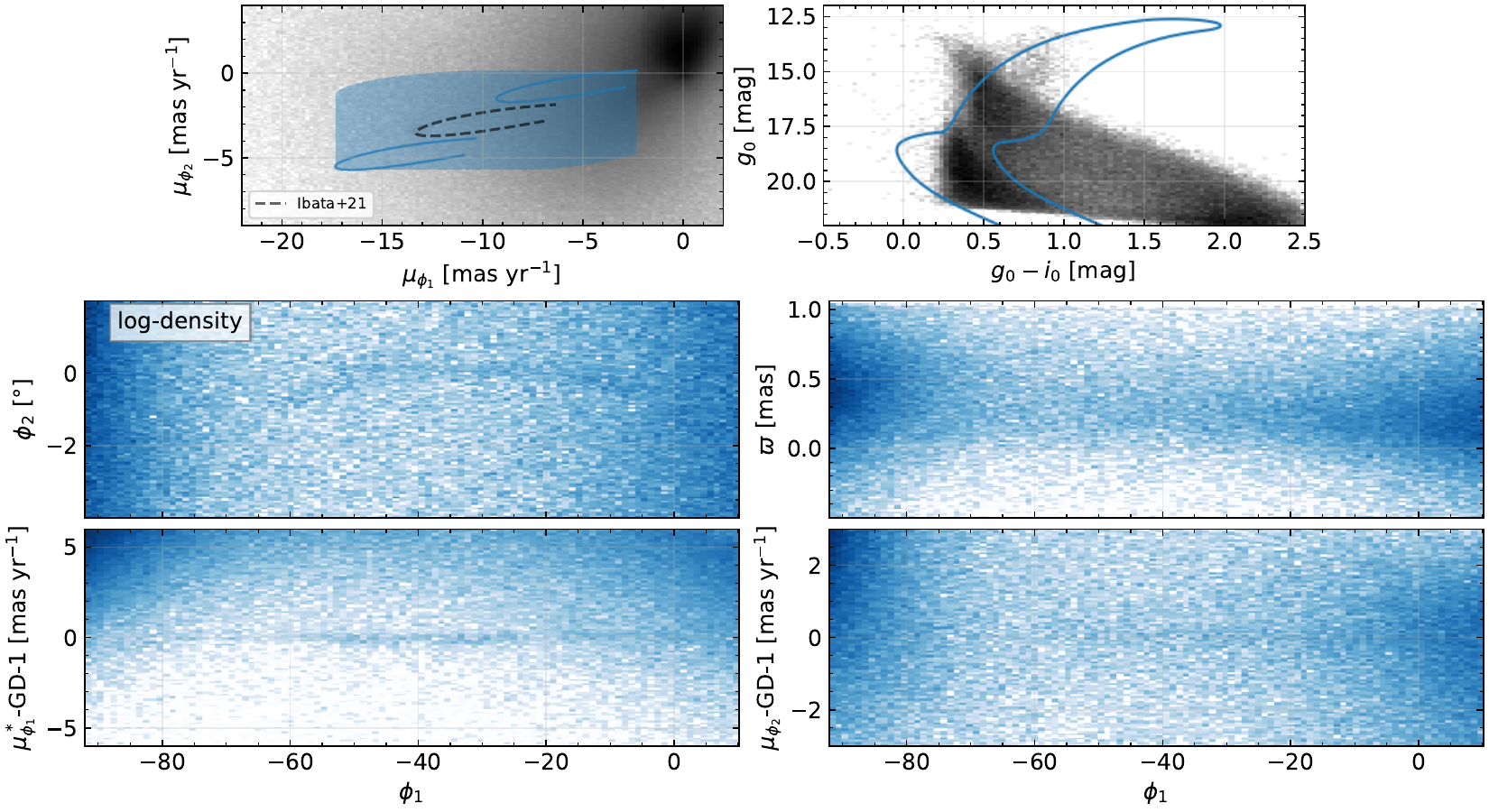}
        \caption{%
            Data selections for \stream{GD-1}, within the data cube described in
            \autoref{sub:results_gd1:data}.
            \textbf{Row 1:}
            Proper motion and photometric selections.  
            The proper motion selection is the $[\mu_{\phi_1},
            \mu_{\phi_2}](\phi_1)$ curve from \citetalias{Ibata+2021} (dashed
            black line) buffered by $6$ and $3$ mas/yr, respectively (blue
            curves defining the bottom left and top-right of the shaded
            rectangular region). Viewed without $\phi_1$ we see the
            $\mu_{\phi_1}, \mu_{\phi_2}$ ``phase''-like diagram. The photometric
            selection is based on a 12 Gyr, [Fe/H] = -1.2 MIST isochrone,
            buffered by 0.3 mag to select the stream. The data is cut at
            $G\leq20$, before applying the isochrone selection.
            \textbf{Rows 2 \& 3:}
            Applying the combination of astrometric and photometric selections,
            plotted for astrometric coordinates $\phi_2, \parallax,
            \mu_{\phi_1}, \mu_{\phi_2}$.  The proper motions are plotted
            relative to \citetalias{Ibata+2021}. Using a log-density coloring,
            the stream is identifiable in all phase-space coordinates except
            $\parallax$ where the errors are significant.  }
        \label{fig:gd1-data_selection}
    \end{figure*}

        The result of applying our density model to the synthetic data is
        illustrated in \autoref{fig:mock_data_result}. The top panel shows the
        weight parameter that controls membership probability as a function of
        $\phi_1$. The dashed curve represents the output of our data-driven
        model, while the histogram is the ground truth. The second panel shows
        the stream and background in the $(\phi_1,\phi_2)$ plane, and the third
        panel shows the stream and background in the $(\phi_1,\parallax)$ plane.
        Points are color-coded according to their membership probability
        \autoref{eq:membership_prob}. The weight parameter clearly shows that
        the model has captured the two prominent density variations along the
        stream.  There are additional variations, mostly around the edges of the
        stream, though re-sampling the neural network weights with dropout
        reveals that these variations represent regions of higher model
        uncertainty. 

        We find that the stream is successfully recovered in both astrometric
        and photometric coordinates, with
        $>99\%$ 
        of stream stars being recovered with membership probability greater than
        $80\%$.  In the bottom two rows of \autoref{fig:mock_data_result} we
        illustrate the synthetic data color-coded by stream membership
        probability (top) and background membership probability (background) for
        different $\phi_1$ slices. Importantly, the photometric portion of the
        mixture model successfully identifies the stream in magnitude space,
        showing a high stream membership probability where the stream is
        present, and a low probability ($\sim 0$) where the stream is absent.
        The false positive rate (i.e., stars misidentified as stream members)
        for this test is found to be $<0.1\%$ (3 stars) when cutting on the 80\%
        membership probability threshold.


\section{Results: GD-1} \label{sec:results_gd1}

    We now apply our method for stream characterization to the stellar stream
    \stream{GD-1}, initially discovered by \citet{GrillmairDionatos2006}.
    Previous work has identified several density variations along \stream{GD-1}
    \citep{deBoer+2018}, including a bifurcation that can be modeled by an
    encounter with a dense dark matter subhalo \citep{Bonaca+2019, Webb+2019}.
    Characterizing the detailed density fluctuations of this stream is therefore
    important for constraining a population of perturbers.  We discuss the data
    selection in \autoref{sub:results_gd1:data}, model specifics in
    \autoref{sub:results_gd1:model}, and the results in
    \autoref{sub:results_gd1:results}.

    \subsection{Data Selection} \label{sub:results_gd1:data}

        We utilize data from \Gaia{} DR3 for the astrometric coordinates of each
        star.  To obtain more accurate color information, we cross-match the
        \Gaia{} field on \PanStarrs{} \citep{PanSTARRS1} using \Gaia's provided
        \texttt{Best-neighbors} catalog.  To limit the field to a region known
        to contain \stream{GD-1} we adopt broad cuts on the cross-matched data.
        We perform these rough cuts by requiring the data lie within the
        phase-space cube:
        \begin{itemize}
            \setlength\itemsep{0em}
            \item $\phi_1 \in [-100, 40] \ \rm{deg}$,
            \item $\phi_2 \in [-9, 5] \ \rm{deg}$,
            \item $\parallax \in [-10, 1] \ \rm{mas}$,
            \item $G_{\rm BP} \in [-1, 3] \ \rm{mag}$
            \item $g, r \in [0, 50] \ \rm{mag}$, 
        \end{itemize}
        where we use \package{gala}'s \citep{gala, galav1.3} \code{GD1Koposov10}
        frame \citep{Koposov+2010} to define $\phi_1$, $\phi_2$, and associated
        proper motions.  For downloading purposes we split the query into 32
        sub-regions in even $\phi_1$ increments.  Stream-specific coordinates
        ($\phi_1, \phi_2$) are not included in \Gaia.  One could embed the
        coordinate transform in the ADQL \citep{ADQL2.0}, or define great-circle
        arcs in ICRS \citep{ICRS1997}. Since only rough cuts are required and
        since the sub-region boxes are sufficiently small for a reasonable
        flat-sky approximation  we simply translate the ($\phi_1, \phi_2$) boxes
        to ICRS.  We correct the \PanStarrs{} photometry for dust extinction
        using the \textit{Bayestar19} \citep{Green+2019} models in the
        \package{dustmaps} package \citep{Green2018}.  \textit{Bayestar19} is a
        three-dimensional dust map of the Milky Way for which we use \Gaia's
        position and parallax distance information. We note that the extinction
        has very little dependence on the distance past $\approx 8 \ \rm{kpc}$.
        For stars with missing parallax measurements we assume a distance of
        $8.5 \ \rm{kpc}$, a typical distance to the stream, thus increasing
        potential membership probability. Adopting a slightly lower or higher
        fiducial distance does not significantly change the results of our
        analysis. We include the 1$\sigma$ errors as a source of uncertainty in
        $\mbs{\Sigma}_n$, the per-star photometric uncertainty.  We also correct
        the parallax for the zero-point using the
        \href{https://pypi.org/project/gaiadr3-zeropoint/}{\code{gaiadr3-zeropoint}}
        package, which implements the findings from \cite{Lindegren+2021}.  The
        result is a field populated by %
  8,243,417\label{output/gd1/ndata_variable.txt}\unskip%

        stars, from which we would like to identify stream members.

        We create data masks for a higher signal-to-noise selection around
        \stream{GD-1}.  The two mask types, proper motion and photometric, are
        shown in \autoref{fig:gd1-data_selection}. For the proper motions we use
        \citet{Ibata+2021} (hereafter \citetalias{Ibata+2021}) as the reference
        for the known track of the stream. We select all all stars with
        $\mu_{\phi_1}^*, \mu_{\phi_2}$ within 6 mas/yr of the track, as a
        function of $\phi_1$. In \autoref{fig:gd1-data_selection} the proper
        motion data are shown as the difference from the \citetalias{Ibata+2021}
        track.  In photometric coordinates we define a mask around a $12 \
        \rm{Gyr}$, [Fe/H] = $-1.2$ isochrone at a distance of $7.8 \ \rm{kpc}$.
        The selection region is a $0.3 \ \rm{mag}$ orthogonal buffer around the
        MIST isochrone \citep[using][]{brutus}, which is illustrated in the
        2nd-from-the-top-left plot in \autoref{fig:gd1-data_selection}. The
        buffered isochrone selection is wide enough to contain \stream{GD-1}
        over its full distance gradient, including the ``typical" distance of
        8.5 kpc used in the photometric correction.  In addition, to avoid
        detailed modeling of the completeness as a term in $\prior_{obs}$ (from
        \autoref{eq:isochrone_prior_components}) we mask out photometry with $G
        > 20$.  With combinations of these masks the stream has much larger
        signal-to-noise, though is still only a small percent of the total data.

    \subsection{Model Specification}\label{sub:results_gd1:model}

        \begin{figure}
            \script{gd1/plot/photometric_background_selection.py}
            \centering
            \includegraphics[width=1\linewidth]{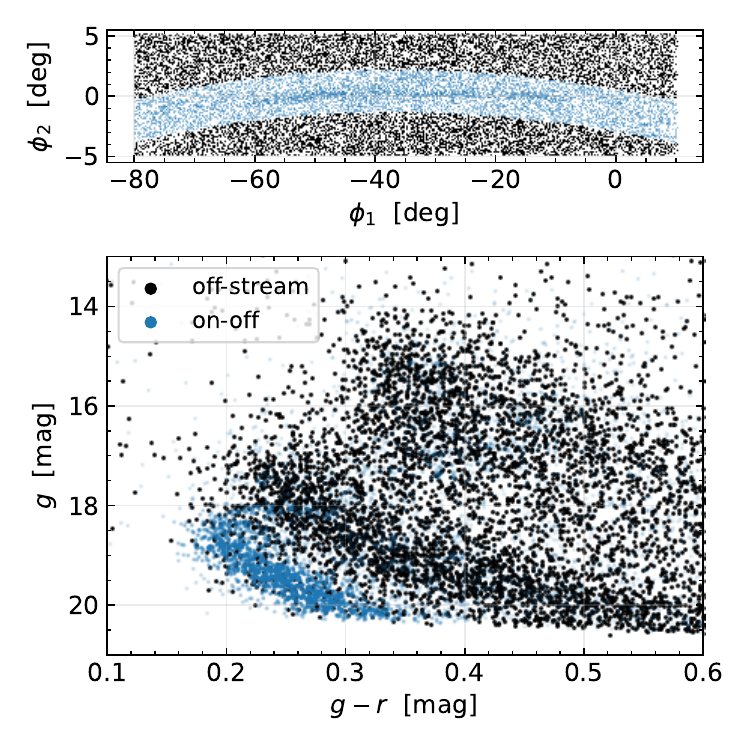}
            \vspace{-20pt}
            \caption{%
                The on and off-stream selection.
                \textbf{Top panel:} Astrometric coordinates ($\phi_1$,
                $\phi_2$).  Light blue points are the stream and background
                within an on-stream region. Black points are in the off-stream
                region used to train the background photometric model. Note this
                plot is generated from a reduced field with a tighter proper
                motion cut to aid in visually identifying the stream.  The
                proper motion cut is $-15 < \mu_{\phi_1}^* < -10$, $-4.5 <
                \mu_{\phi_2} < -2$ [mas/yr].
                \textbf{Bottom panel:} CMD of the astrometric selection above.
                The stream's isochrone is apparent as an overdensity.
            }
            \label{fig:gd1-photometric_background_selection}
        \end{figure}

        The modeling framework, detailed in section \autoref{sec:method}, is
        modular, allowing many model components to be combined, nested, added to
        the mixture, and linked via priors.  The model we require for
        \stream{GD-1} is similar to the model built for the mock stream in
        \autoref{sub:results_mock:model}, but has additional components for
        \stream{GD-1}'s spur.  In this section we will discuss how the total
        model is built, noting similarities to the example mocks-stream model.

        All full stream models necessarily include a background model, which we
        model in both astrometric and photometric spaces.  The photometric
        background is too complex to capture with simple analytic distributions.
        Instead, we train a normalizing flow on the magnitudes $g, r$,
        conditioned on $\phi_1$. We refer readers to
        \autoref{sub:method:pre-training_distributions} for discussion of the
        photometric distribution and how a normalizing flow can characterize
        this complex distribution, including its variation over $\phi_1$.  The
        flow's flexibility, while necessary in this context, also poses a
        problem: the flow captures the stream as easily as the background.
        Consequently we select on-stream and off-stream regions using the known
        $\phi_1,\phi_2$ range of the stream, as shown in the upper panel of
        \autoref{fig:gd1-photometric_background_selection}.  In the lower panel
        we plotted this selection in photometric coordinates, where an isochrone
        is apparent in the on-stream selection.  The normalizing flow is trained
        on the off-stream selection, and will be fixed (not trainable, nor
        contribute to the gradient) when incorporated into the larger model.

        Unlike for the photometry, the astrometric background could be (mostly)
        fit with simple analytic distributions.  This is true if the selected
        field is large enough -- the distribution in each coordinate is
        approximately a skew-normal or a similarly tailed exponential-type
        function. However by defining a higher signal-to-noise field around the
        \citet{Ibata+2021} stream track in $\mu_{\phi_1}^*, \mu_{\phi_2}$, the
        background in these coordinates is no longer well described by an
        exponential distribution. Similarly, for much of the extent of
        \stream{GD1's} field the parallax is well modeled by a truncated
        skew-normal distribution, $\parallax \sim \rm{TruncSkew}\mcal{N}(\mu,
        \Sigma; a, b)$, however for $\phi_1 \in [-100, -60]$ this does not hold.
        Rather than splitting the model at a somewhat arbitrary $\phi_1$ or
        interpolating between two distributions we instead fit the astrometry
        with a $\phi_1$-conditioned normalizing flow. The background model is
        characterized by:

        \begin{itemize}
            \setlength\itemsep{0em}
            \item $\phi_2, \sim \rm{TruncExp}(\lambda; a, b)$: a truncated
                exponential distribution (see
                \autoref{app:sub:exponential_distribution}).  The distribution
                has 1 slope parameter.
            \item $\mu_{\phi_1}^*, \mu_{\phi_2}, \varpi \sim q_{\rm flow}$: a 3
                argument , one context feature ($\phi_1$) normalizing flow
                trained on an off-stream selection (see
                \autoref{fig:gd1-photometric_background_selection} for that
                selection). Though discussed in the context of a photometric
                background, \autoref{sub:method:pre-training_distributions}
                details how these flows are constructed and trained.
        \end{itemize}

        We model the main stream of \stream{GD-1} similarly to the mock stream
        -- as an independent addition of both astrometric and photometric
        models.  The astrometric model for the stream is a 5-dimensional
        Gaussian in a Gaussian track in $\phi_2, \parallax, \mu_{\phi_1},
        \mu_{\phi_2}, \parallax$, using the model defined in
        \autoref{ssub:method:astrometric_model:on_stream}.  The parameter
        network's 10 outputs are $\mbs{\theta} = [\mu_{\phi_2}, \sigma_{\phi_2},
        \mu_\parallax, \sigma_\parallax, \mu_{\mu_{\phi_1}^*},
        \sigma_{\mu_{\phi_1}^*}, \mu_{\mu_{\phi_2}}, \sigma_{\mu_{\phi_2}},
        \mu_\parallax, \sigma_\parallax](\phi_1)$.  In testing the network is
        sufficiently large to permit 15\% dropout and retain robust predictions.

        The stream's photometric model is a distance modulus-shifted isochrone,
        using the model from \autoref{ssub:method:photometric_model:on_stream}.
        We use a $12 \ \rm{Gyr}$, [Fe/H]$ = -1.2$ MIST isochrone with no assumed
        intrinsic dispersion.  \stream{GD-1}'s true spectroscopic metallicity is
        [Fe/H]$\lesssim-2$, but there is enough degeneracy in age and
        metallicity that the chosen isochrone empirically matches the CMD locus
        and is useful in this modeling context.  Using the method described in
        \autoref{ssub:method:linking_to_astrometrics} the distance modulus track
        is set by the parallax track in the astrometric model. Thus the
        on-stream photometric model has no independent parameters and
        consequently no neural network.

  \begin{table}
\centering
\setlength{\tabcolsep}{0pt}
\newcommand\capitem{\\$\phantom{+}\ast$\ }
\caption{%
    Stream Track Regions Prior: %
    This table includes all the region priors used to guide the model towards
    the known stream track. The model will converge to the region $_{\rm
    minimum}^{\rm maximum}$.
    See \autoref{sub:methods:priors:track_region_prior} for details.
}
\label{tab:gd1_track_prior}
{\rowcolors{2}{gray!10}{white!10}
\begin{tabular}{@{}r<{\hspace{7pt}}*{4}{r<{\hspace{7pt}}}l<{\hspace{7pt}}@{}}
\toprule
component & $\phi_1 \,[\rm{\degree}]$ & $\phi_2 \,[\rm{\degree}]$ & $\mu_{\phi_1} \,[\frac{\rm{mas}}{\rm{yr}}]$ & $\mu \,[\rm{mag}]$ & $(\simeq \parallax \,\unit{mas})$ \\
\midrule
stream & -90.0 & ${\color{gray}\big(}_{-5.28}^{-1.78}{\color{gray}\big)}$ &  &  &  \\
stream & -80.0 & ${\color{gray}\big(}_{-3.93}^{-0.43}{\color{gray}\big)}$ &  &  &  \\
stream & -70.0 & ${\color{gray}\big(}_{-2.89}^{\phantom{+}0.61}{\color{gray}\big)}$ & ${\color{gray}\big(}_{-12.20}^{-10.20}{\color{gray}\big)}$ & ${\color{gray}\big(}_{14.20}^{14.80}{\color{gray}\big)}$ & ${\color{gray}\big(}_{0.11}^{0.15}{\color{gray}\big)}$ \\
stream & -60.0 & ${\color{gray}\big(}_{-2.13}^{\phantom{+}1.37}{\color{gray}\big)}$ & ${\color{gray}\big(}_{-13.30}^{-11.70}{\color{gray}\big)}$ &  &  \\
stream & -50.0 & ${\color{gray}\big(}_{-1.66}^{\phantom{+}1.84}{\color{gray}\big)}$ & ${\color{gray}\big(}_{-14.05}^{-12.55}{\color{gray}\big)}$ &  &  \\
stream & -40.0 & ${\color{gray}\big(}_{-1.46}^{\phantom{+}2.04}{\color{gray}\big)}$ & ${\color{gray}\big(}_{-14.05}^{-12.55}{\color{gray}\big)}$ & ${\color{gray}\big(}_{14.10}^{14.70}{\color{gray}\big)}$ & ${\color{gray}\big(}_{0.11}^{0.15}{\color{gray}\big)}$ \\
spur & -35.0 & ${\color{gray}\big(}_{0.45}^{2.15}{\color{gray}\big)}$ & ${\color{gray}\big(}_{-14.95}^{-10.95}{\color{gray}\big)}$ &  &  \\
stream & -30.0 & ${\color{gray}\big(}_{-1.51}^{\phantom{+}1.99}{\color{gray}\big)}$ & ${\color{gray}\big(}_{-13.35}^{-11.85}{\color{gray}\big)}$ &  &  \\
spur & -30.0 & ${\color{gray}\big(}_{0.45}^{2.15}{\color{gray}\big)}$ & ${\color{gray}\big(}_{-14.60}^{-10.60}{\color{gray}\big)}$ &  &  \\
stream & -20.0 & ${\color{gray}\big(}_{-1.80}^{\phantom{+}1.70}{\color{gray}\big)}$ &  &  &  \\
spur & -20.0 & ${\color{gray}\big(}_{0.65}^{2.35}{\color{gray}\big)}$ & ${\color{gray}\big(}_{-13.50}^{\phantom{+}-9.50}{\color{gray}\big)}$ &  &  \\
stream & -10.0 & ${\color{gray}\big(}_{-2.32}^{\phantom{+}1.18}{\color{gray}\big)}$ & ${\color{gray}\big(}_{-11.75}^{\phantom{+}-8.25}{\color{gray}\big)}$ &  &  \\
stream & 0.0 & ${\color{gray}\big(}_{-3.05}^{\phantom{+}0.45}{\color{gray}\big)}$ &  & ${\color{gray}\big(}_{14.70}^{15.30}{\color{gray}\big)}$ & ${\color{gray}\big(}_{0.08}^{0.12}{\color{gray}\big)}$ \\
stream & 5.0 &  & ${\color{gray}\big(}_{-9.15}^{-5.65}{\color{gray}\big)}$ &  &  \\
stream & 10.0 & ${\color{gray}\big(}_{-3.97}^{-0.47}{\color{gray}\big)}$ &  &  &  \\
\bottomrule\bottomrule
\end{tabular}
}
\end{table}
\label{output/gd1/control_points.tex}\unskip%

        We model the spur of \stream{GD-1} as a separate component to the main
        stream. The spur has an identical model as the stream: a 5-D astrometric
        Gaussian and isochrone in the photometry. Owing to their common origin,
        we impose the prior that the spur and stream share the same isochrone.
        The parallax is also shared, motivated by initial findings, and that
        there were too few stars to robustly detect any deviations.

  \begin{table*}[htp]
\centering
\small
\setlength{\tabcolsep}{0pt}
\newcommand\capitem{\\$\phantom{+}\ast$\ }

\caption{Subset of GD-1 Membership Table.
\\
This table includes a selection of candidate member stars for the GD-1 stream,
based on the membership likelihoods.  For each star we include the Gaia DR3
source ID and astrometric solution, the Pan-STARRS1 photometry, and the
membership likelihoods for the stream, spur, and background.  The likelihoods
are computed using the trained model described in
\autoref{sub:results_gd1:results} and we include a quality flag ${\rm
dim}(\boldsymbol{x})$, indicating the number of features used by the model.  For
most stars all features are measured.  We use dropout regularization to estimate
the uncertainty in the likelihoods, and report the 5\% and 95\% quantiles of the
distribution, as well as the dropout-disabled maximum-likelihood estimate (MLE)
of the likelihood.
\\
We include as interesting cases:
    \capitem{} 1 star with the highest MLE for the stream,
    \capitem{} 5 stars with high stream MLE ($\mathcal{L}^{(S)}_{\rm MLE} > 0.9$),
    \capitem{} 4 stars with low stream MLE, but whose 95\% likelihood is high
               ($\mathcal{L}^{(S)}_{\rm MLE} < 0.75, \mathcal{L}^{(S)}_{\rm 95\%} > 0.8$),
    \capitem{} 1 star with the maximum MLE for the spur,
    \capitem{} 1 star with high spur MLE and low stream MLE
               ($\mathcal{L}^{(spur)}_{\rm MLE} > 0.9, \mathcal{L}^{(S)}_{\rm MLE} < 0.75$),
\\
For convenience we round the likelihoods to 2 decimal places, and only show the
value and uncertainty when it is non-zero.
\\
\textit{The full table, including source ids, is available online.}
}
\label{tab:gd1_member_table}
\begin{tabular}{@{}c<{\hspace{7pt}}*{5}{>{\footnotesize}c<{\hspace{7pt}}}*{2}{>{\footnotesize}c<{\hspace{7pt}}}c<{\hspace{7pt}}*{3}{l<{\hspace{7pt}}}@{}}
\toprule
\multicolumn{6}{c}{Gaia} & \multicolumn{2}{c}{PS-1} & \multicolumn{1}{c}{} & \multicolumn{3}{c}{Membership Likelihood (${\rm MLE}_{\phantom{0}5\%}^{95\%}$)}\\
\cmidrule(lr){1-6} \cmidrule(lr){7-8} \cmidrule(lr){10-12}
\texttt{source\_id} & $\alpha$ [$\mathrm{{}^{\circ}}$] & $\delta$ [$\mathrm{{}^{\circ}}$] & $\mu_{\alpha}^{*}$ [$\frac{\rm{mas}}{\rm{yr}}$] & $\mu_{\delta}$ [$\frac{\rm{mas}}{\rm{yr}}$] & $\varpi$ [\rm{mas}] & g [mag] & r [mag] & ${\rm dim}(\boldsymbol{x})$ & $\mathcal{L}_{\rm stream}$ & $\mathcal{L}_{\rm spur}$ & $\mathcal{L}_{\rm background}$ \\
\midrule
--- & 207.32 & 58.53 & $-7.71 \pm 0.01$ & $-2.85 \pm 0.01$ & $0.08 \pm 0.01$ & $14.14 \pm 0.01$ & $13.37 \pm nan$ & 7 & $1.00$ & --- & --- \\
\rowcolor{gray!7}
--- & 138.13 & 20.99 & $-3.78 \pm 0.27$ & $-12.66 \pm 0.20$ & $0.22 \pm 0.28$ & $19.15 \pm 0.02$ & $18.91 \pm 0.02$ & 7 & $0.99_{-0.04}^{+0.00}$ & --- & $0.01_{-0.00}^{+0.04}$ \\
--- & 166.42 & 49.40 & $-7.08 \pm 0.06$ & $-9.72 \pm 0.08$ & $0.04 \pm 0.09$ & $17.53 \pm 0.00$ & $17.14 \pm 0.00$ & 7 & $0.99_{-0.10}^{+0.01}$ & --- & $0.01_{-0.01}^{+0.10}$ \\
--- & 159.37 & 45.40 & $-6.94 \pm 0.22$ & $-11.34 \pm 0.22$ & $0.43 \pm 0.27$ & $19.76 \pm 0.02$ & $19.48 \pm 0.01$ & 7 & $0.96_{-0.42}^{+0.03}$ & $0.00_{-0.00}^{+0.03}$ & $0.04_{-0.02}^{+0.41}$ \\
--- & 146.05 & 32.95 & $-5.13 \pm 0.25$ & $-12.72 \pm 0.18$ & $0.36 \pm 0.23$ & $19.02 \pm 0.02$ & $18.81 \pm 0.02$ & 7 & $0.95_{-0.11}^{+0.02}$ & $0.03_{-0.02}^{+0.08}$ & $0.01_{-0.00}^{+0.04}$ \\
--- & 145.70 & 32.85 & $-5.14 \pm 0.38$ & $-12.48 \pm 0.26$ & $0.18 \pm 0.40$ & $19.74 \pm 0.02$ & $19.47 \pm 0.02$ & 7 & $0.95_{-0.13}^{+0.02}$ & $0.04_{-0.02}^{+0.13}$ & $0.01_{-0.01}^{+0.01}$ \\
--- & 190.27 & 57.04 & $-7.52 \pm 0.43$ & $-5.50 \pm 0.40$ & $0.16 \pm 0.38$ & --- & --- & 5 & $0.74_{-0.61}^{+0.12}$ & --- & $0.26_{-0.12}^{+0.61}$ \\
\rowcolor{gray!7}
--- & 135.37 & 16.22 & $-2.82 \pm 0.79$ & $-12.77 \pm 0.52$ & $0.72 \pm 0.88$ & --- & --- & 5 & $0.72_{-0.43}^{+0.13}$ & --- & $0.28_{-0.13}^{+0.43}$ \\
\rowcolor{gray!7}
--- & 164.08 & 47.92 & $-8.06 \pm 0.16$ & $-10.93 \pm 0.20$ & $0.19 \pm 0.21$ & $19.28 \pm 0.01$ & $19.07 \pm 0.01$ & 7 & $0.69_{-0.66}^{+0.16}$ & --- & $0.31_{-0.16}^{+0.66}$ \\
\rowcolor{gray!7}
--- & 146.51 & 34.16 & $-5.01 \pm 0.25$ & $-13.06 \pm 0.22$ & $0.33 \pm 0.23$ & $19.07 \pm 0.01$ & $18.85 \pm 0.01$ & 7 & $0.63_{-0.30}^{+0.18}$ & $0.30_{-0.18}^{+0.23}$ & $0.08_{-0.02}^{+0.16}$ \\
\rowcolor{gray!7}
--- & 156.99 & 44.64 & $-6.41 \pm 0.31$ & $-11.47 \pm 0.40$ & $0.74 \pm 0.40$ & $19.86 \pm 0.06$ & $19.63 \pm 0.05$ & 7 & --- & $0.99_{-0.01}^{+0.00}$ & $0.01_{-0.00}^{+0.01}$ \\
--- & 149.88 & 38.71 & $-5.87 \pm 0.25$ & $-12.42 \pm 0.20$ & $0.00 \pm 0.25$ & $19.34 \pm 0.02$ & $19.12 \pm 0.01$ & 7 & $0.00_{-0.00}^{+0.01}$ & $0.96_{-0.12}^{+0.01}$ & $0.04_{-0.01}^{+0.12}$ \\
\bottomrule\bottomrule
\end{tabular}
\end{table*}
\label{output/gd1/member_table_select.tex}\unskip%

        The parameter phase-space is large and a random initialization will take
        significant time to converge towards the stream track. This initial
        training is unnecessary as the rough stream track is known \textit{a
        priori}.  To guide the astrometric model towards the known
        \stream{GD-1}, we set down guides (see
        \autoref{sub:methods:priors:track_region_prior}) along \stream{GD-1}'s
        main track and on the spur.  We set the width of these guides to large
        values, so that their influence on the model is not dominant. Indeed, we
        find that the value of the guides is to help reduce training time, since
        otherwise it will take the flexible density model much longer to
        converge to the actual stream.  We include all the track regions in
        \autoref{tab:gd1_track_prior}.

        All the component models are tied together into a mixture model.  The
        singular weight parameter network is a \texttt{LTD} network with 15\%
        dropout.  The stream weights are in the range $\ln f \in [-10, 0]$,
        while the spur weights are $\ln f \in [-10, 0]$ and set to 0 for $\phi_1
        < -45, -15 < \phi_1$.  By \autoref{eq:mixture_weight_normalization} the
        background brings the cumulative weight to one, normalizing the mixture.

    \subsection{Trained Density Model}\label{sub:results_gd1:results}

        \begin{figure*}[ht]
            \script{gd1/plot/results_full.py}
            \centering
            \includegraphics[width=0.83\linewidth]{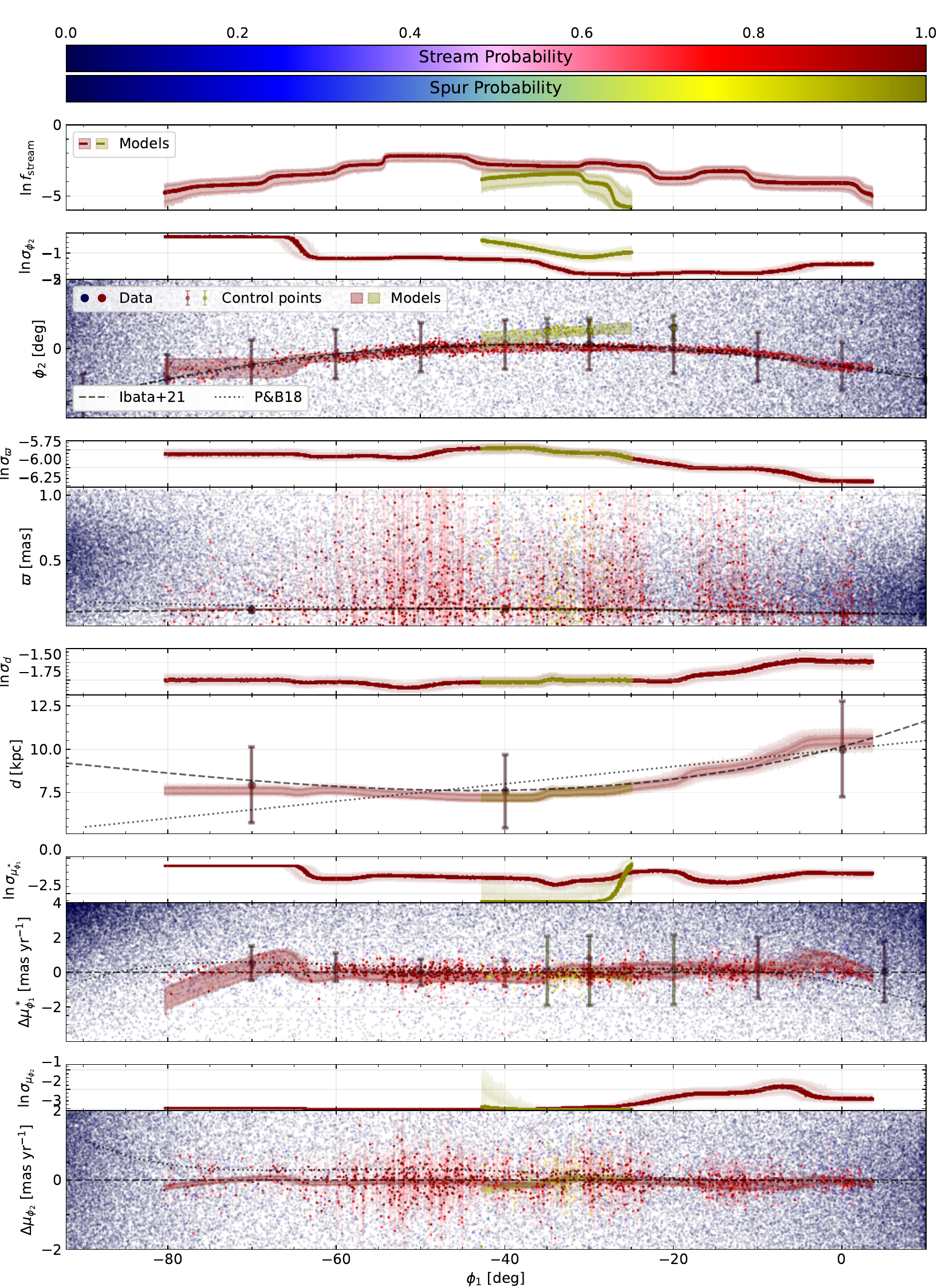}
            \caption{%
                The model for \stream{GD-1}, including both a thin-stream and
                spur component.  For comparison we include the tracks from
                \citet{Price-WhelanBonaca2018} and \citet{Ibata+2021}. Note how
                the model largely agrees with the latter track in all
                astrometric dimensions.
                \textbf{Panel 1: }%
                    The stream mixture coefficients $f(\phi_1)$ predicted by the
                    model, colored by the stream and spur membership likelihood.
                \textbf{Panel 2: }
                    $\phi_2(\phi_1)$ over the full range of $\phi_1$. The data
                    color and transparency are set by the model's
                    50th-percentile stream and spur probability. We include the
                    error bars for stars with $>75\%$ membership probability.
                    The predicted 50th-percentile track $\pm$ width is
                    over-plotted (red band), along with the 90\% confidence
                    region (broader red band).  The 5th to 95th percentile
                    variation in the width is shown  in the adjoining top plot.
                \textbf{Panel 3 \& 4: }%
                    $\parallax(\phi_1)$ and $d(\phi_1)$ over the full range of
                    $\phi_1$. The $\parallax$ is tied to the photometry -- see
                    \autoref{fig:gd1-results-panels}.
                \textbf{Panel 5 \& 6: }%
                    The proper motions $\Delta \mu_{\phi_1^*}, \Delta
                    \mu_{\phi_2}$ shown relative to the galstreams track (black
                    dashed line). The model convolves observational errors with
                    the stream width, but the large fractional errors make the
                    determination difficult.   }
            \label{fig:gd1-results-full}
        \end{figure*}

        \begin{figure*}[t]
            \script{gd1/plot/results_panels.py}
            \centering
            \includegraphics[width=1\linewidth]{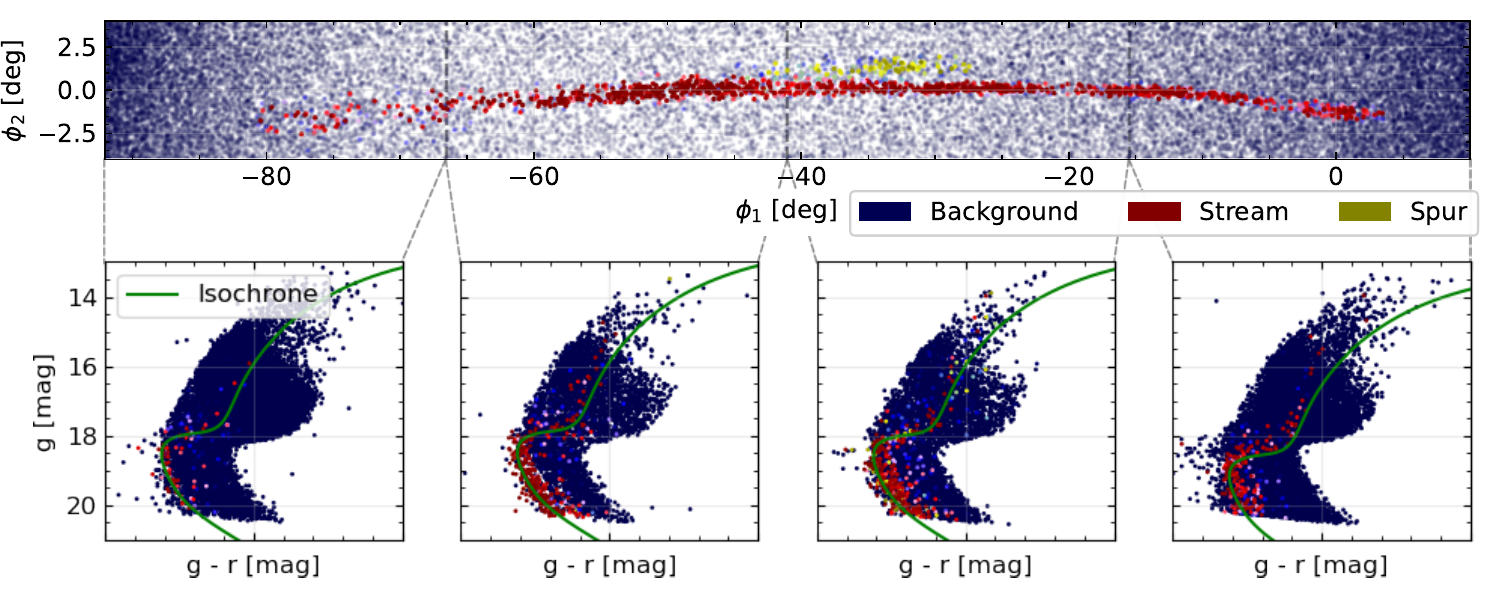}
            \caption{%
                Photometric coordinate $(g, r)$ plots for each model component
                in 4 $\phi_1$ ranges across the data set. Over-plotted is the
                model isochrone, shifted to the mean distance of the track
                within the selected $\phi_1$ range. Importantly this means we do
                not expect the mean isochrone to exactly match the distribution
                of points. See \autoref{fig:gd1-results-full} for the match of
                the distance gradient to the data and prior literature results.
            }
            \label{fig:gd1-results-panels}
        \end{figure*}

\begin{figure}[ht]
            \script{gd1/plot/smooth_likelihood.py}
            \centering
            \hspace{-25 pt}\includegraphics[width=1.05\linewidth]{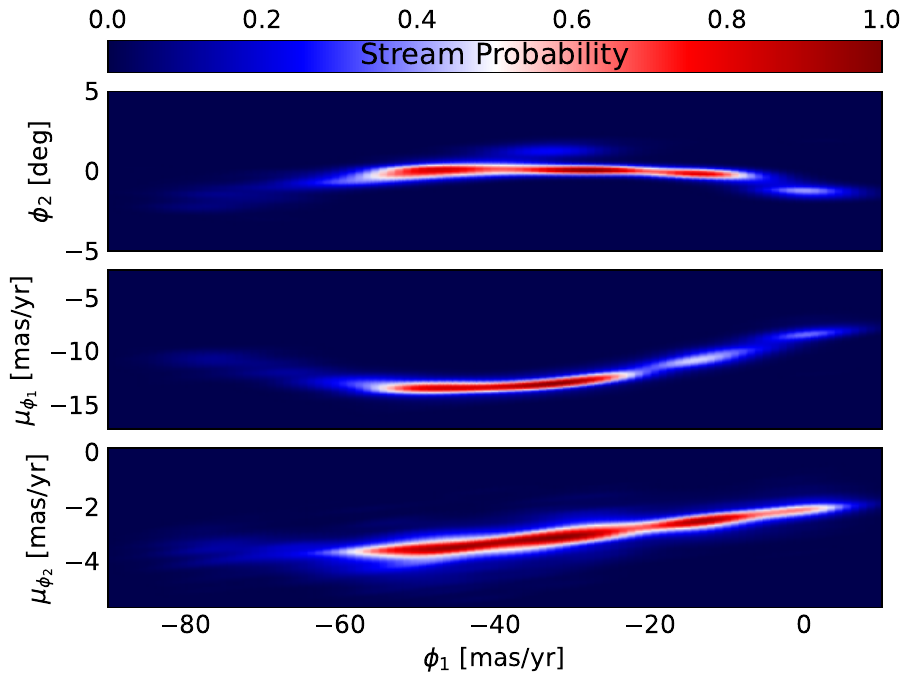}
            \caption{%
                Smoothed KDE model of \stream{GD-1}'s member stars, weighted by
                their membership probability. The KDE has a bandwidth of $0.1 \
                \rm{deg}$ (or mas/yr).  In $\phi_2$ the spur is clearly visible
                in the KDE.  The gaps at $-20\degree$ are prominent in both
                $\phi_2$ and $\mu_{\phi_1}$.  The gap at $-40\degree$ is most
                prominent in $\phi_2$.
            }
            \label{fig:gd1-heatmap}
        \end{figure}

        In this section we present results from our model applied to real {\it
        Gaia} and Pan-STARRS data of \stream{GD-1}.  We visualize the
        performance of our model in identifying stream stars in
        \autoref{fig:gd1-results-full}, and illustrate our photometric fit in
        \autoref{fig:gd1-results-panels}. In \autoref{tab:gd1_member_table} we
        provide a selection of \stream{GD-1} member stars. The full catalog is
        available online at
        \href{https://zenodo.org/records/10211410}{zenodo:10211410}. Lastly,
        \autoref{fig:gd1-heatmap} is a membership probability-weighted density
        heatmap of the \stream{GD-1} stream derived from our fit.

        In \autoref{fig:gd1-results-full}, the top panels shows the fraction of
        stars belonging to the stream (red) and spur (yellow) as a function of
        $\phi_1$. The second panel illustrates the stream in $\phi_1$, $\phi_2$
        coordinates, color-coded by the membership probability for the stream
        and spur components (derived from \autoref{eq:membership_prob}). The
        following rows represent the other astrometric dimensions for the stream
        and spur components, with the distance track obtained through our
        photometric model (visualized in \autoref{fig:gd1-results-panels} and
        discussed below). The red and yellow bands represent model predictions
        for the main stream and spur, respectively. The points with thick error
        bars represent control points (discussed in
        \autoref{sub:results_gd1:model}) for the stream and spur components,
        which are priors on the location of the stream and spur components
        across the astrometric dimensions. The control points in the parallax
        panel represent the reliable parallax and distance modulus estimates
        from \citet{deBoer+2020}. The actual data points are color-coded by the
        maximum membership probability of belonging to any of the three model
        components (i.e., maximum of the main stream, spur, or background
        components). For stars with membership probability greater than $75\%$,
        we also plot their astrometric errorbars from \Gaia{} as the thin red
        lines.

        Because the neural networks are trained with dropout, we can estimate
        the modeling uncertainty in our fits to each stream with a Monte Carlo
        procedure \citep{GalGhahramani2015}. Namely, by incorporating dropout
        during both neural network training and inference, the distribution of
        model outputs represents the posterior predictive distribution,
        marginalized over the neural network parameters. An estimate of this
        distribution is shown in the weight parameter of
        \autoref{fig:gd1-results-full}, represented as the red (main stream) and
        yellow (spur) bands. For the astrometric dimensions, our model predicts
        a mean track, and a standard deviation. Error bands are estimated by
        incorporating dropout with a rate of 15\%. Each of the panels in
        \autoref{fig:gd1-results-full} shows the track mean and width,
        calculated from the 50th percentile of the posterior predictive
        distribution for both quantities. Plotted behind that is the 90\%
        credible region of the track mean. We note that the credible region of
        the mean alone is generally larger than that of the $\mu \pm \sigma$
        50th-percentile ridge. This highlights the importance of exploring the
        full posterior of parameters, not only a Maximum Likelihood Estimation
        (MLE) or posterior ridge-line.

        Our model finds %
  $730$\label{output/gd1/nstream/nstream_variable.txt}\unskip%
stars
        that belong to the main stream component with membership probability
        greater than
  $80\%$\label{output/gd1/nstream/minimum_membership_probability.txt}\unskip%
 at the
  $50$th\label{output/gd1/nstream/posterior_percentile.txt}\unskip%
 percentile in the
        membership posterior distribution. For the spur component, there are
  $37$\label{output/gd1/nspur/nspur_variable.txt}\unskip%
 stars with the same
        membership probability threshold at the same percentile of the posterior
        distribution.  A tabulation of the astrometry, photometry, and
        membership posterior distribution for each star is included in
        \autoref{tab:gd1_member_table}. The full table may be found on the
        paper's
        \href{https://github.com/nstarman/stellar_stream_density_ml_paper}{GitHub
        repository}. \autoref{tab:gd1_member_table} gives the MLE prediction,
        but also samples the 5th-95th percentile of the posterior distribution
        of the likelihood using the dropout method. The variation in the 90\%
        credible region in the track means there can be significant variation in
        a star's estimated membership probability. 

        The extent of the main stream in $\phi_1$ is roughly $80~\rm{deg}$, and
        the angular extent of the spur is roughly $15~\rm{deg}$. The distance
        track (bottom panel) places the stream in the range of $6-10~\rm{kpc}$,
        with an upwards concavity as a function of $\phi_1$. This result is
        consistent with other works, which find a similar distance to
        \stream{GD-1} using other techniques \citep{deBoer+2020, Ibata+2021}.
        Importantly, our distance estimate is not hard-coded to fall in this
        range: we allow for the stream track to fall anywhere within
        $[4,40]~\rm{kpc}$, with a broad prior around previous literature
        estimates. The similarity of our inferred distance track compared to
        other works indicates that our photometric model is performing
        satisfactorily.

        We also recover the well-known underdensity gaps along the stream around
        $\phi_1 \approx -40, -20 \ \rm{and } -5~\rm{deg}$ (seen best in
        \autoref{fig:gd1-heatmap}). The $-20\degree$ feature was first reported
        by \citet{CarlbergGrillmair2013}, and is one location hypothesized for
        the dissolved stream progenitor. \citet{WebbBovy2019} suggests the
        $-40\degree$ gap might be  the progenitor location.  The recovery of
        this density fluctuation from a joint model in astrometric and
        photometric coordinates developed here argues that the feature is indeed
        intrinsic to the stream and not an artifact of the data.

        The spur is a distinct component of our model, independent in its weight
        and all astrometric dimensions except $\parallax$ which is shared with
        the \stream{GD-1} main stream through the photometric model. The model
        finds a non-negligible spur feature, with a peak fractional weight
        approaching the thin stream. As corroboration, the proper motion of the
        spur closely agrees with the thin stream. The alignment of these
        dimensions is not hard-coded in our model. The width of the spur in
        proper motion is very thin, thinner even than the stream.  However with
        the large observational errors and low number density of the spur it is
        difficult to attribute this to a physical difference between the two
        systems. In contrast, the small errors in $\phi_2$ indicate the spur's
        width diverges from the thin stream at $\phi_1 > -35 \ \rm{deg}$,
        whereas the width is similar at lower $\phi_1$. There are numerous
        mechanisms by which a stream's width may vary \citep[e.g.  epicyclic
        effects as in][]{Ibata+2020}, so differences between the spur and stream
        are both notable and expected. 

        \stream{GD-1} has other observed tidal features, namely the ``blob''
        \citep{Price-WhelanBonaca2018}, which is a very diffuse component
        enclosing the thin stream. When fitting our model to \stream{GD-1} with
        a more stringent proper motion cut, we are able to recover the blob. In
        a more crowded field, the stream makes up only a few percent of the
        total number of stream stars, so diffuse features like the blob are not
        separated from the background. We could, perhaps, guide the model
        towards the blob by adding an additional mixture distribution that is
        specialized to diffuse components around the thin stream. Because the
        aim of the current paper is to introduce and demonstrate our method, we
        defer a more detailed fit of the diffuse extra-tidal structure of Milky
        Way streams to future work.

        We note that the model-predicted track does not always closely align
        with the predicted member stars or our notion of a slowly-evolving
        stream. This misalignment is most evident in $\mu_{\phi_1^*}$, where the
        track has a sharp turn up at $\phi_1 \sim -65 \ \rm{deg}$ and then
        flattens out. There are a number of contributing factors. First, the
        observational uncertainties are large enough that this flattening is
        consistent with the data. Second, we do not enforce the stream track to
        be smooth with $\phi_1$ -- our fits are dictated by the data.  Last,
        examining 2D slices when fitting in high dimensional spaces can be
        misleading \citep{Aggarwal+2002}. Unsurprisingly, the raw membership
        probabilities for each model component provide a more accurate
        representation of the stream's 6d phase-space distribution than any
        phase-space slice could, as may be seen in \autoref{fig:gd1-heatmap}. We
        discuss the smoothed stream model determined solely by membership
        probabilities at the end of this section.
        
        We highlight the performance of our photometric model in
        \autoref{fig:gd1-results-panels}. In this figure, we color-code each
        star by its membership probability (top row; same as
        \autoref{fig:gd1-results-full}), and in the bottom row we visualize the
        performance of our photometric model in bands of $\phi_1$. Our model
        isochrone, which shifts vertically due to our distance modulus estimate,
        is shown in green (we plot the mean-distance isochrone in each $\phi_1$
        band). Because our model preforms a joint fit in photometry and
        astrometry, the most probable stream members do not need to fall exactly
        along the isochrone.  Still, the likely stream members are
        overwhelmingly distributed as might be expected from a single-stellar
        population.  Even the more diffuse tails of the stream can be seen in
        the CMD distribution, where they still roughly follow a single-stellar
        population. 
        
        With membership probabilities for each component of the stream estimated
        from both astrometry and photometry, it is possible to quickly construct
        a custom representation of the stream's phase-space density. We
        illustrate this in  \autoref{fig:gd1-heatmap}, which shows a Gaussian
        kernel density estimate (KDE) of bandwidth $0.1 \ \rm{deg}$ (or mas/yr)
        fit to the model. The primary overdensities at $\phi_1 = (-50, -30, -10)
        \ \rm{deg}$ are all visible in $\phi_2$. The gap at $20 \ \rm{deg}$ is
        prominent in all features, but the gap at $-40$ is less visible in
        proper motions, since the spur shares the stream's proper motions and
        thus contributes to the density in those coordinates. This highlights
        the importance of constructing a multidimensional smooth density model
        of streams.

\section{Results: Palomar 5} \label{sec:results_pal5}

    We now apply the method to the stellar stream \stream{Palomar 5}.
    Historically lengths of the stream have been found with matched filter
    photometric techniques, like with the discovery in \citet{Odenkirchen+2001}.
    Post-\Gaia{} the stream was also observed by filtering kinematically
    \cite[e.g.,][]{Starkman+2019, Ibata+2021}.  However at $\sim 20 \ \rm{kpc}$,
    the \stream{Pal\,5} main sequence is not observed by \Gaia{}, making a high
    signal-to-noise astrometric detection challenging.  We discuss the data
    selection in \autoref{sub:results_pal5:data}, model specifics in
    \autoref{sub:results_pal5:model}, and the results in
    \autoref{sub:results_pal5:results}.

    \begin{figure*}[ht]
        \script{pal5/plot/data_selection.py}
        \centering
        \includegraphics[width=1\linewidth]{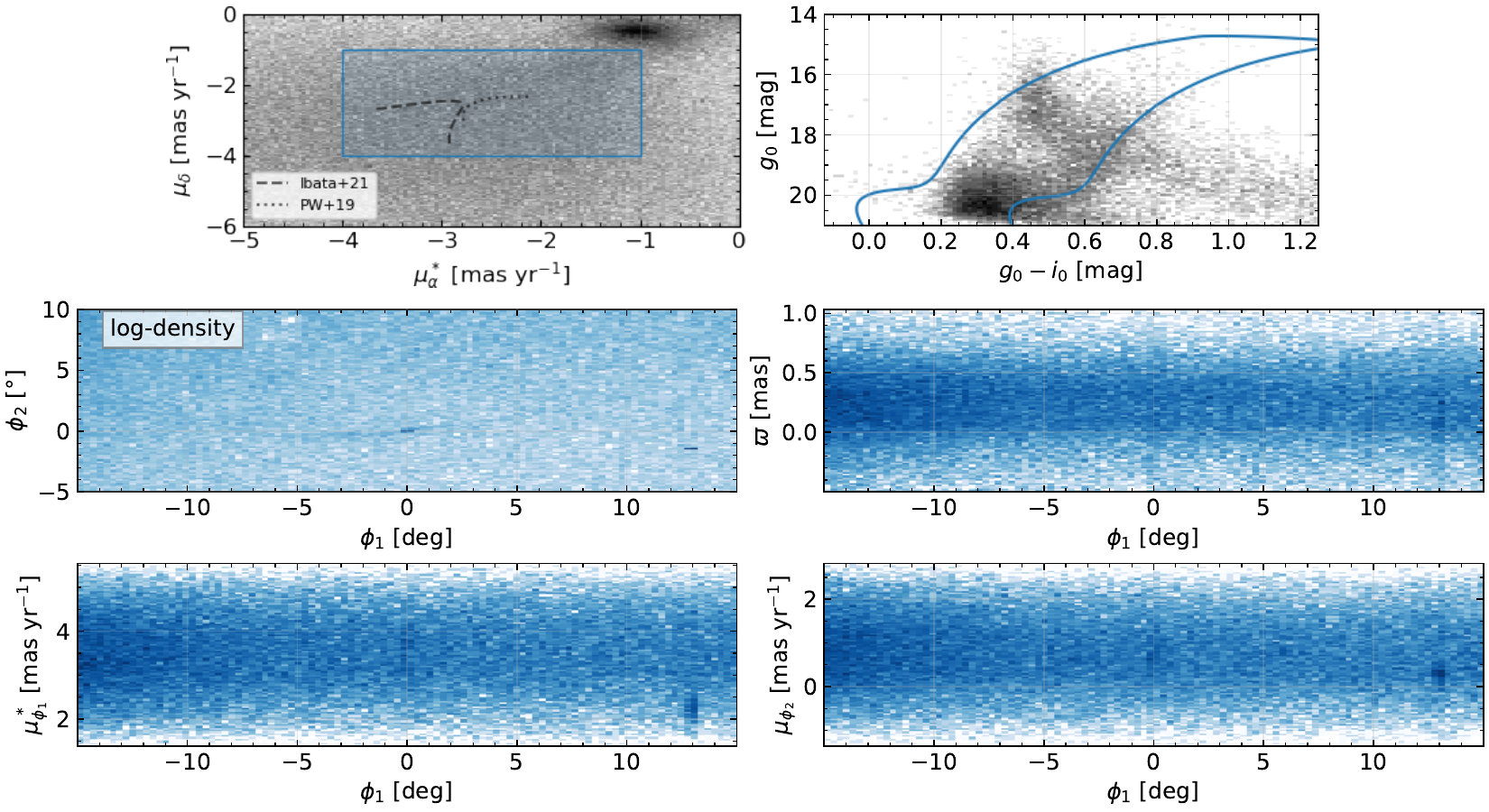}
        \caption{%
            Data selections for \stream{Pal\,5}, within the data cube described
            in \autoref{sub:results_pal5:data}.
            \textbf{Row 1:}
            Proper motion and photometric selections.  \stream{Pal\,5} is
            apparent in the proper motions as a small overdensity. The
            photometric selection is based on a 12 Gyr, [Fe/H] = -1.2 MIST
            isochrone, buffered by 0.3 dex. The photometry is also cut to
            \Gaia{}'s $G\leq20$ mag, before applying the isochrone selection.
            \textbf{Rows 2 \& 3:}
            Applying the combination of astrometric and photometric selections,
            plotted for astrometric coordinates $\phi_2, \parallax,
            \mu_{\phi_1}, \mu_{\phi_2}$.  The stream is an identifiable
            overdensity in only $\phi_2$ and errors dominate all other
            coordinates.  }
        \label{fig:pal5-data_selection}
    \end{figure*}

    \subsection{Data Selection} \label{sub:results_pal5:data}

        We construct a \Gaia{} DR3 - \PanStarrs{} crossmatched field very
        similarly to the procedure outlined in \autoref{sub:results_gd1:data}
        for \stream{GD-1}.  To limit the field to a region containing
        \stream{Pal\;5} we adopt broad cuts on the data by requiring the data
        lie within the phase-space cube:
        \begin{itemize}
            \setlength\itemsep{0em}
            \item $\phi_1 \in [-20, 20] \ \rm{deg}$,
            \item $\phi_2 \in [-5, 10] \ \rm{deg}$,
            \item $\parallax \in [-10, 1] \ \rm{mas}$,
            \item $\rm{G-R} \in [-0.5, 1.2] \ \rm{mag}$
            \item $g, r \in [0, 30] \ \rm{mag}$,
        \end{itemize}
        where we use \package{gala}'s \code{Pal5PriceWhelan18} frame
        \citep{Price-Whelan+2019} to define $\phi_1$, $\phi_2$, and associated
        proper motions.  Using the \textit{Bayestar19} maps we correct the
        \PanStarrs{} photometry for dust extinction.  For stars without distance
        information we assume a distance of $20 \ \rm{kpc}$, a typical distance
        to the stream \citep{Harris1996, Odenkirchen2003, Grillmair2006,
        Ibata+2016, Price-Whelan+2019} and thus increasing the probability of
        the star being considered a member.  The 1$\sigma$ dust correction
        errors are included by quadrature in $\mbs{\Sigma}_n$, the per-star
        photometric uncertainty. We also correct the parallax for the \Gaia\
        zero-point as in \autoref{sub:results_gd1:data}.  The result is a field
        populated by %
  5,022,757\label{output/pal5/ndata_variable.txt}\unskip%
 stars, from which
        we would like to identify stream members.

        Within the broad selection cuts, we create more specific data masks.  In
        particular we mask concentrated over-densities, like the globular
        cluster \stream{M5}, that are not associated with \stream{Pal\;5}.  In
        \autoref{fig:pal5-data_selection} we show the proper motion and
        photometric masks, with \stream{M5} already filtered out.  The proper
        motion mask extends from $\mu_{\phi_1}$ and $\mu_{\phi_2} \in (-4, -1)$
        [mas/yr].  The photometric mask is defined by a $0.15 \ \rm{mag}$ buffer
        around a $[\rm{Fe}/{H}] = -1.3$, $11.5 \ \rm{Gyr}$ isochrone. It is the
        shape of the isochrone which is important, not the physical reality of
        the metallicity nor age, discussed in further detail in
        \autoref{ssub:method:photometric_model:on_stream}.  In addition, to
        avoid detailed modeling of the completeness as a term in $\prior_{obs}$
        (from \autoref{eq:isochrone_prior_components}) we mask out photometry
        with $G > 20$.  With combinations of these masks the stream has much
        larger signal-to-noise, though is still only a small fraction of the
        total data.

        The reduced dataset, with all specific masks applied, is included below
        the data mask plots in \autoref{fig:pal5-data_selection}. In $\phi_2$,
        $\mu_{\phi_1}^*$, and $\mu_{\phi_2}$ the \stream{Pal\,5} progenitor is
        clearly visible. Only in $\phi_2$ is the stream visible, and then only
        for a few degrees around the progenitor. The distance of \stream{Pal\,5}
        means most $\parallax$ measurements have $\sim 100\%$ errors, so neither
        the stream nor even the progenitor are evident.


    \vspace{-6pt}
    \subsection{Model Specification} \label{sub:results_pal5:model}

        In this section we discuss how the total \stream{Pal\,5} model is built,
        using the modular modeling framework from \autoref{sec:method}.  We
        model the stream of \stream{Pal\;5} as a single Gaussian component
        conditioned on $\phi_1$. As a function of $\phi_1$, the astrometric
        stream density is a 3-dimensional Gaussian in $\phi_2, \mu_{\phi_1},
        \mu_{\phi_2}$, using the model defined in
        \autoref{ssub:method:astrometric_model:on_stream}.  The parameter
        network is a 5-layer \texttt{LTD} where the 6 outputs are $\mbs{\theta}
        = [\mu_{\phi_2}, \sigma_{\phi_2}, \mu_{\mu_{\phi_1}^*},
        \sigma_{\mu_{\phi_1}^*}, \mu_{\mu_{\phi_2}},
        \sigma_{\mu_{\phi_2}}](\phi_1)$.  In testing, the network is
        sufficiently large to permit 15\% dropout and retain robust predictions.
        At $\sim 20 \ \rm{kpc}$ any \Gaia{}-crossmatched data is not
        photometrically deep enough to contain the main sequence nor its turn
        off. Thus we do not model the stream photometrically. We only use
        photometric cuts to remove some background. Similarly, we do not include
        the parallax as the errors are too large for the parallax to contribute
        to the model. 

        \begin{figure}[ht]
            \script{pal5/plot/photometric_background_selection.py}
            \centering
            \includegraphics[width=1\linewidth]{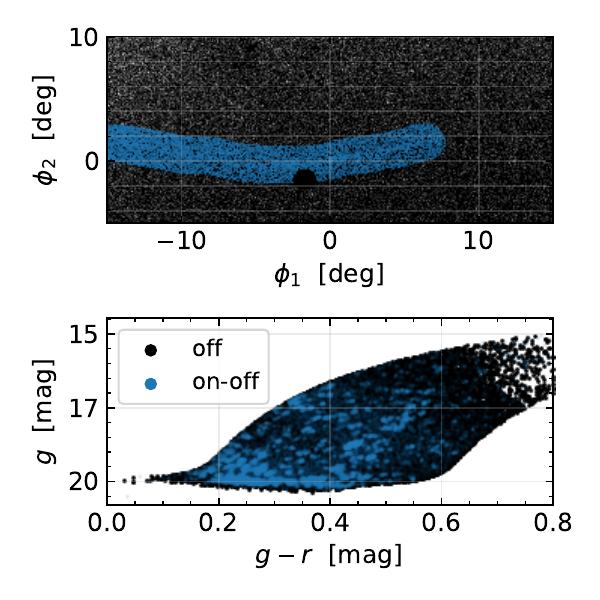}
            \caption{%
                \textbf{Top panel:} Astrometric coordinates ($\phi_1$,
                $\phi_2$).  Light blue points are the stream and background
                within an ``on"-stream region. Black points are in the
                ``off"-stream region used to train the background photometric
                model. The \stream{M5} cluster is masked as a black circle
                cutting into the on-stream selection at $\phi_1\sim-1\degree$.
                \textbf{Bottom panel:} CMD plot of the astrometric selection
                above. The stream's isochrone is somewhat apparent, but the
                background contamination is large.
            }
            \label{fig:pal5-photometric_background_selection}
        \end{figure}
        
        To avoid a large convergence time, we use as a prior the known stream
        track, in this case \package{galstream}'s \citep{Mateu2022}
        implementation of \stream{Pal\,5} from \citet{Ibata+2021}. From the
        track, we set down guides (see
        \autoref{sub:methods:priors:track_region_prior}) regularly spaced in
        $\phi_1$. We include all the guides in \autoref{tab:pal5_track_prior}.
        We don't place guides in the other coordinates, except at the location
        of the progenitor, because the stream is not visible and the models
        disagree on the proper motions (see \autoref{fig:pal5-data_selection},
        top left plot).  The progenitor is an exception, both visible
        kinematically and also extremely well measured
        \citep{VasilievBaumgardt2021}, and we place a tight kinematic prior on
        the stream track at the location of the progenitor.  The guide priors
        are visible in \autoref{fig:pal5-results-full}.
  \begin{table}[htp]
\centering
\setlength{\tabcolsep}{0pt}
\newcommand\capitem{\\$\phantom{+}\ast$\ }
\caption{%
    Stream Track Regions Priors for Pal\,5: %
    This table includes all the region priors used to guide the model towards
    the known stream track. The model will converge to the region $_{\rm
    minimum}^{\rm maximum}$. The regions are determined by the stream track
    from \package{galstreams} \citep{Mateu2022}, with a width significantly
    larger than the stream width. In the kinematics only the progenitor is used
    to guide the model.
}
\label{tab:pal5_track_prior}
{\rowcolors{2}{gray!10}{white!10}
\begin{tabular}{@{}*{4}{r<{\hspace{7pt}}}@{}}
\toprule
$\phi_1 \,[\rm{\degree}]$ & $\phi_2 \,[\rm{\degree}]$ & $\mu_{\phi_1} \,[\frac{\rm{mas}}{\rm{yr}}]$ & $\mu_{\phi_2} \,[\frac{\rm{mas}}{\rm{yr}}]$ \\
\midrule
-12.71 & ${\color{gray}\big(}_{-0.08}^{\phantom{+}1.92}{\color{gray}\big)}$ &  &  \\
-10.30 & ${\color{gray}\big(}_{-0.44}^{\phantom{+}1.56}{\color{gray}\big)}$ &  &  \\
-7.89 & ${\color{gray}\big(}_{-0.78}^{\phantom{+}1.22}{\color{gray}\big)}$ &  &  \\
-5.48 & ${\color{gray}\big(}_{-1.05}^{\phantom{+}0.95}{\color{gray}\big)}$ &  &  \\
-3.07 & ${\color{gray}\big(}_{-1.19}^{\phantom{+}0.81}{\color{gray}\big)}$ &  &  \\
-0.66 & ${\color{gray}\big(}_{-1.15}^{\phantom{+}0.85}{\color{gray}\big)}$ &  &  \\
0.00 & ${\color{gray}\big(}_{-0.10}^{\phantom{+}0.10}{\color{gray}\big)}$ & ${\color{gray}\big(}_{2.96}^{4.24}{\color{gray}\big)}$ & ${\color{gray}\big(}_{0.09}^{1.37}{\color{gray}\big)}$ \\
1.76 & ${\color{gray}\big(}_{-0.88}^{\phantom{+}1.12}{\color{gray}\big)}$ &  &  \\
4.17 & ${\color{gray}\big(}_{-0.32}^{\phantom{+}1.68}{\color{gray}\big)}$ &  &  \\
6.58 & ${\color{gray}\big(}_{0.58}^{2.58}{\color{gray}\big)}$ &  &  \\
8.99 & ${\color{gray}\big(}_{1.88}^{3.88}{\color{gray}\big)}$ &  &  \\
\bottomrule\bottomrule
\end{tabular}
}
\end{table}
\label{output/pal5/control_points.tex}\unskip%

        The background is fit with an analytic distribution and a normalizing
        flow:
        \begin{itemize}
            \setlength\itemsep{0em}
            \item $\phi_2 \sim \rm{TruncExp}(\lambda; a, b)$: a truncated
                exponential distribution (see
                \autoref{app:sub:exponential_distribution}).  The distribution
                has 1 parameter and we use a \texttt{LTD} neural network with
                15\% dropout.
            \item $\mu_{\phi_1}, \mu_{\phi_2} \sim q_{\rm flow}$: a two feature,
                one context feature normalizing flow trained on an off-stream
                selection (see
                \autoref{fig:pal5-photometric_background_selection} for that
                selection). \autoref{sub:method:pre-training_distributions}
                details how these flows are constructed and trained.
        \end{itemize}

        All the component models are tied together into a mixture model.  The
        weight parameter network is the usual \texttt{LTD} with 15\% dropout.
        The stream weights are in the range $\ln{f_{\rm stream}} \in [-10, 0]$
        and set to $-\infty$ for $\phi_1 \notin [-10^\circ, 10^\circ]$. By
        \autoref{eq:mixture_weight_normalization} the background  brings the
        cumulative weight to one, normalizing the mixture.


    \subsection{Trained Density Model} \label{sub:results_pal5:results}

        \begin{figure*}[ht]
            \script{pal5/plot/results_full.py}
            \centering
            \includegraphics[width=0.85\linewidth]{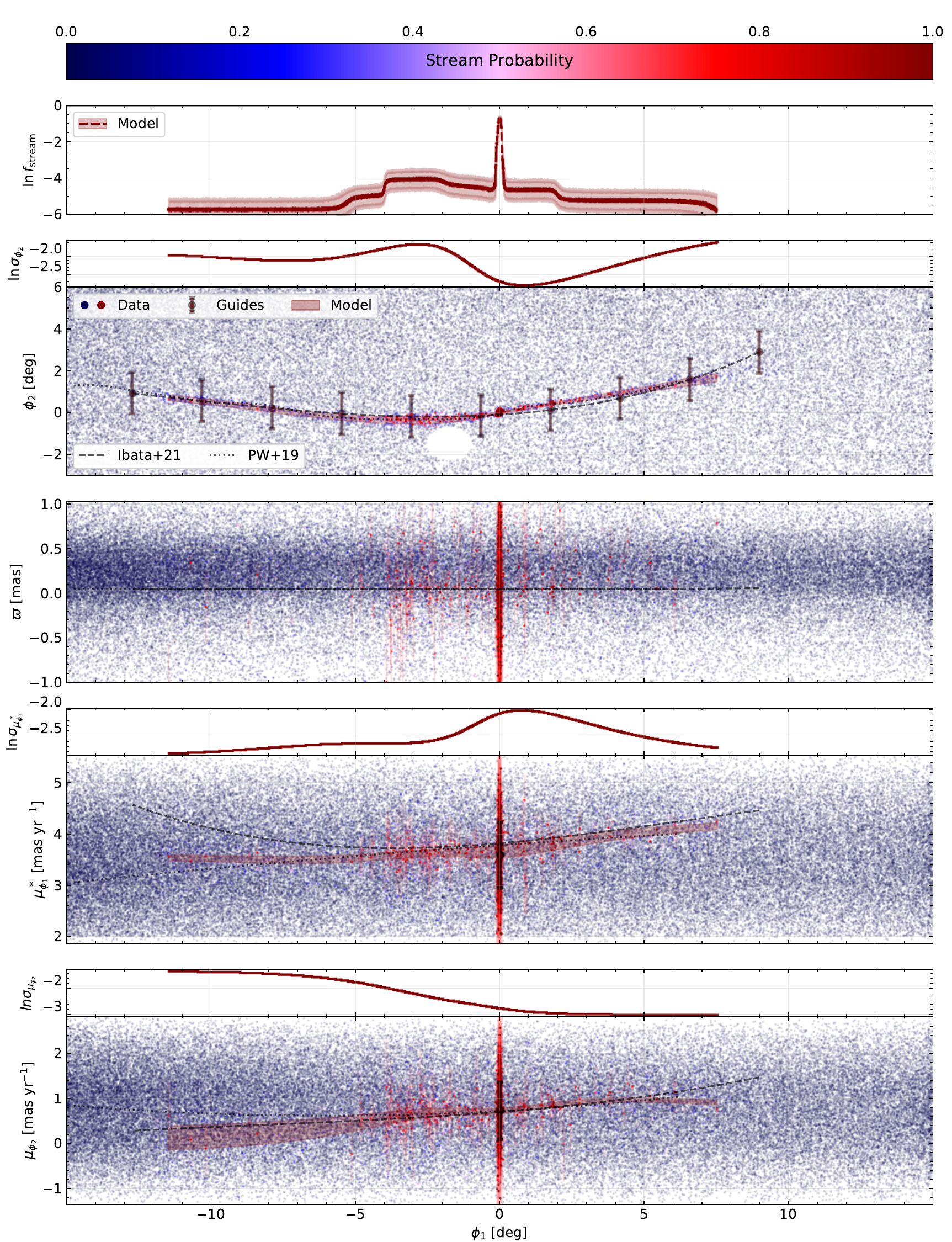}
            \caption{%
                The model for \stream{Pal\,5}.  For comparison we include the
                tracks from \citet{Price-Whelan+2019} and \citet{Ibata+2021}.
                \textbf{Panel 1: }%
                The stream mixture coefficients $f(\phi_1)$ predicted by the
                model, colored by the stream membership likelihood. The
                stream fraction and number density drop to nearly $0$ for $\phi_1 > 5 \
                \rm{deg}$, making the track dominated by low-number
                statistics.
                \textbf{Panel 2: }
                $\phi_2(\phi_1)$ over the full range of $\phi_1$. The data color
                and transparency are set by the model's 50th-percentile stream
                probability. We include the error bars for stars with $>75\%$
                membership probability. The predicted 50th-percentile track
                $\pm$ width is over-plotted (red band), along with the 90\%
                confidence region (broader red bend). The 5th to 95th percentile
                variation in the width is shown in the adjoining top plot.
                \textbf{Panel 3: }%
                $\parallax(\phi_1)$ and $d(\phi_1)$ over the full range of
                $\phi_1$.
                \textbf{Panel 4 \& 5: }%
                The proper motions $\mu_{\phi_1^*}, \mu_{\phi_2}$. The model
                convolves observational errors with the stream width, but the
                large fractional errors make the determination difficult. The
                track is largely consistent with 0 width driven by small number
                statistics.  }
            \label{fig:pal5-results-full}
        \end{figure*}

        \begin{figure*}[htp]
            \script{pal5/plot/results_panels.py}
            \centering
            \includegraphics[width=1\linewidth]{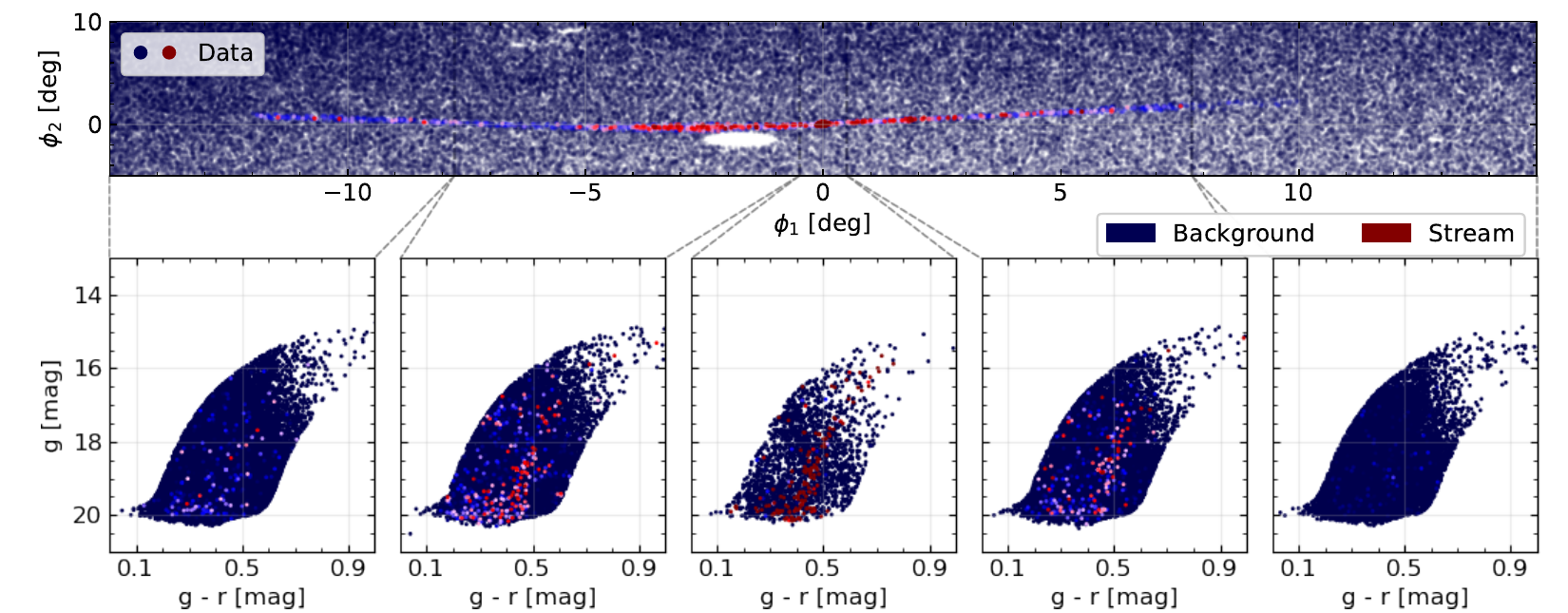}
            \caption{%
                Photometric coordinate $(g, r)$ plots for each model component
                in 5 $\phi_1$ ranges across the data set, centered on the
                \stream{Pal\,5} progenitor.
                On either side of the progenitor we strongly detect the stream
                for $\approx 5\degree$ and tentatively for an additional
                $\approx 5\degree$. The leading arm (positive $\phi_1$) track
                for $\phi_1 > 5\degree$ is consistent with
                \citet{Starkman+2019}. We do not detect any fanning of the
                stream, but given the shallow photometry, this is expected.
            }
            \label{fig:pal5-results-panels}
        \end{figure*}

  \begin{table*}[htp]
\centering
\small
\setlength{\tabcolsep}{0pt}
\newcommand\capitem{\\$\phantom{+}\ast$\ }

\caption{Subset of Pal\;5 Membership Table.
\\
This table includes a selection of candidate member stars for the Pal\;5 stream,
based on the membership likelihoods.  For each star we include the Gaia DR3
source ID and astrometric solution, the Pan-STARRS1 photometry, and the
membership likelihoods for the stream and background. The likelihoods are
computed using the trained model described in \autoref{sub:results_pal5:results} and we
include a quality flag ${\rm dim}(\boldsymbol{x})$, indicating the number of
features used by the model.  For most stars all features are measured. We use
dropout regularization to estimate the uncertainty in the likelihoods, and
report the 5\% and 95\% quantiles of the distribution, as well as the
dropout-disabled maximum-likelihood estimate (MLE) of the likelihood.
\\
We include as interesting cases:
    \capitem{} 1 star with the highest MLE for the stream,
    \capitem{} 5 stars with high stream MLE ($\mathcal{L}^{(S)}_{\rm MLE} > 0.9$),
    \capitem{} 4 stars with low stream MLE, but whose 95\% likelihood is high
               ($\mathcal{L}^{(S)}_{\rm MLE} < 0.75, \mathcal{L}^{(S)}_{\rm 95\%} > 0.8$),
\\
For convenience we round the likelihoods to 2 decimal places, and only show the
value and uncertainty when it is non-zero.
\\
\textit{The full table, including source ids, is available online.}
}
\begin{tabular}{@{}c<{\hspace{7pt}}*{4}{c<{\hspace{7pt}}}*{2}{c<{\hspace{7pt}}}c<{\hspace{7pt}}*{2}{l<{\hspace{7pt}}}@{}}
\toprule
\multicolumn{5}{c}{Gaia} & \multicolumn{2}{c}{PS-1} & \multicolumn{1}{c}{} & \multicolumn{2}{c}{Membership (${\rm MLE}_{5\%}^{95\%}$)}\\
\cmidrule(lr){1-5} \cmidrule(lr){6-7} \cmidrule(lr){9-10}
\texttt{source\_id} & $\alpha$ [$\mathrm{{}^{\circ}}$] & $\delta$ [$\mathrm{{}^{\circ}}$] & $\mu_{\alpha}^{*}$ [$\frac{\rm{mas}}{\rm{yr}}$] & $\mu_{\delta}$ [$\frac{\rm{mas}}{\rm{yr}}$] & g [mag] & r [mag] & ${\rm dim}(\boldsymbol{x})$ & $\mathcal{L}_{\rm stream}$ & $\mathcal{L}_{\rm background}$ \\
\midrule
--- & 228.98 & -0.10 & $-2.74 \pm 0.04$ & $-2.64 \pm 0.03$ & $15.99 \pm 0.02$ & $15.29 \pm 0.02$ & 6 & $1.00$ & --- \\
\rowcolor{gray!7}
--- & 229.03 & -0.03 & $-2.77 \pm 0.57$ & $-1.92 \pm 0.51$ & $19.92 \pm 0.02$ & $19.55 \pm 0.01$ & 6 & $0.98_{-0.01}^{+0.01}$ & $0.02_{-0.01}^{+0.01}$ \\
--- & 228.97 & -0.10 & $-2.66 \pm 0.74$ & $-1.82 \pm 0.65$ & --- & --- & 4 & $0.99$ & $0.01$ \\
--- & 229.05 & -0.20 & $-3.44 \pm 0.67$ & $-2.77 \pm 0.57$ & $20.05 \pm 0.01$ & $19.67 \pm 0.01$ & 6 & $0.98_{-0.01}^{+0.00}$ & $0.02_{-0.00}^{+0.01}$ \\
--- & 229.00 & -0.09 & $-2.30 \pm 0.59$ & $-2.81 \pm 0.51$ & --- & --- & 4 & $1.00$ & --- \\
--- & 229.03 & -0.13 & $-1.52 \pm 0.86$ & $-3.60 \pm 0.73$ & --- & --- & 4 & $0.98_{-0.01}^{+0.00}$ & $0.02_{-0.00}^{+0.01}$ \\
\bottomrule\bottomrule
\end{tabular}
\end{table*}
\label{output/pal5/member_table_select.tex}\unskip%

        Our fit to \stream{Pal\,5} and its tidal tails is illustrated in
        \autoref{fig:pal5-results-full}. In the top panel we plot the stream
        fraction and its uncertainty, estimated by applying dropout during
        inference and evaluation of our model. The model clearly identifies the
        progenitor cluster $(\phi_1 = 0)$. 
    
        The membership probability of stream stars is illustrated in subsequent
        panels of \autoref{fig:pal5-results-full}, with guide points and their
        widths shown in red. The leading and trailing arms of the stream are
        clearly identified in our fits, with high membership probability on
        either side of the cluster. Our model prefers a thin on-sky morphology
        for \stream{Pal\,5}'s tidal tails, with an angular width of the stream
        (parameterized by $\ln{\sigma_{\phi_2}}$) around $\sim 0.1~\rm{deg}$.
        This is contrary to our fit of \stream{GD-1} in
        \autoref{fig:gd1-results-full}, where we find a more appreciable width
        of $\sim 0.2~\rm{deg}$ and above.  This inferred width is consistent
        with \stream{Pal\,5} being more distant than \stream{GD-1}, though we
        expect a thin inferred width for \stream{Pal\,5} since we do not model
        the stream in photometric coordinates. When photometry is available, we
        expect that our photometric model provides substantial constraining
        power in characterizing the stream, since diffuse parts of the tidal
        tails in on-sky angular coordinates remain tightly clustered in the CMD.
        This clustering is seen in our \stream{GD-1} fits
        (\autoref{fig:gd1-results-panels}), where even the more extended parts
        of the stream remain clustered in the CMD.
        
        The parallax, proper motion tracks, and their widths are also shown in
        \autoref{fig:pal5-results-full}. The red vertical lines represent the
        astrometric errors in each dimension, shown only for stars with
        membership probability greater than $75\%$. The errors are large
        compared to the width of the best fit stream track (shown as the red
        band). The track is a valid minimum to our loss function: because our
        model incorporates the large astrometric uncertainties, small variations
        to the estimated width of the stream in each dimension do not produce
        substantial changes to the loss function. Thus, a narrow stream in the
        kinematic dimensions provides a valid fit to the stream given the data
        at hand. 
    
        We also note that the track of the stream shown as the red band in each
        dimension appears to have an abrupt change in direction around $\phi_1
        \approx 7^\circ$. This also represents the region with the lowest stream
        fraction, as seen in the top panel of \autoref{fig:pal5-results-full}.
        In low stream fraction regions, the model is uncertain about the track
        of the stream since only a few stars contribute to the on-stream fit.
        The track changes highlight the importance of utilizing the membership
        probabilities of individual stars when analyzing any given stream, since
        membership probabilities incorporate information on the stream fraction
        while the mean stream-track alone does not provide this information. 
    
        In \autoref{fig:pal5-results-panels} we again show the data in
        $\phi_1-\phi_2$ coordinates, color-coded by membership probabilities in
        the top panel (same colorbar as in \autoref{fig:pal5-results-full}). In
        the bottom panels we plot the CMD of the data in bands of $\phi_1$. As
        discussed, our fit to \stream{Pal\,5} does not incorporate information
        from the CMD.  Thus, the bottom row represents a data-driven isochrone
        for \stream{Pal\,5} and its tidal tails, obtained by fitting the
        stream's astrometric overdensity. Indeed, the highest probability
        regions of the CMD appear qualitatively similar to a standard isochrone,
        with an extended and clustered sequence. The main sequence appears to be
        missing, which is expected given our cut at $G = 20~\rm{mag}$, below
        which the main sequence of \stream{Pal\,5} resides \citep{Bonaca+2020}.
        The similarity of the color-magnitude distribution of \stream{Pal\,5}'s
        tidal tails to the red end of a standard isochrone highlights the
        validity of our astrometric membership model, since a photometric fit
        was never performed. 

\begin{figure}[ht]
            \script{pal5/plot/smooth_likelihood.py}
            \centering
            \hspace{-25pt}\includegraphics[width=1.05\linewidth]{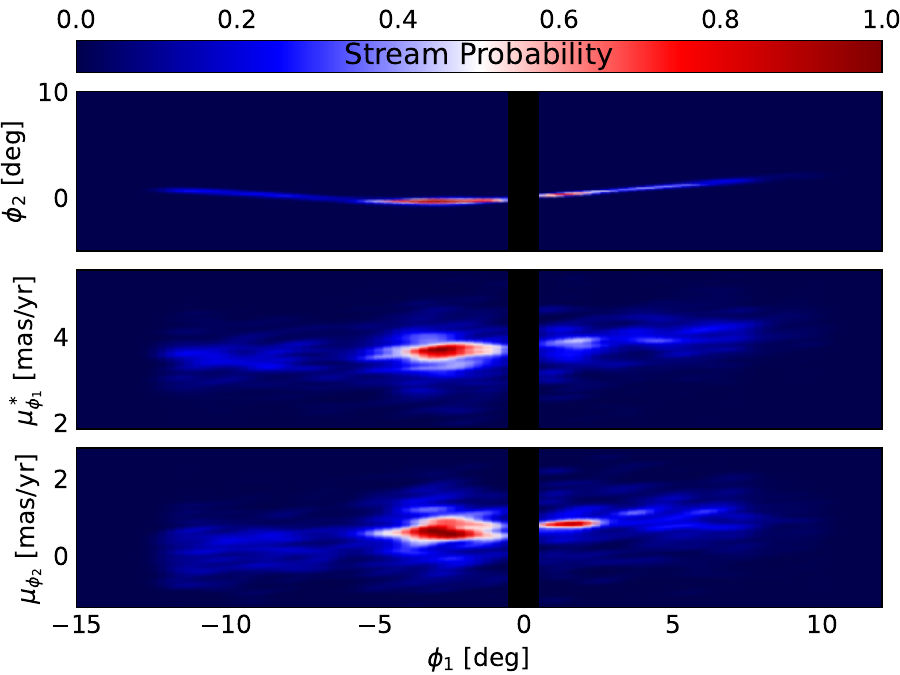}
            \caption{%
                Smoothed KDE model of the stream's member stars, weighted by
                their membership probability. The KDE has a bandwidth of $0.1 \
                \rm{deg}$ (or mas/yr). We exclude the progenitor to limit the
                density range and better display the diffuse tidal tails. The
                stream extends from $-12$ to $10$ degrees in this plot, which
                appears longer than in \autoref{fig:pal5-results-full} and
                \autoref{fig:pal5-results-panels}. Those show the membership
                likelihood at the $50$th-percentile of its distribution. This
                figure samples over the full posterior, showing the contribution
                of less-likely members.
            }
            \label{fig:pal5-heatmap}
        \end{figure}

        The stream solution, i.e. the per-star membership likelihood, is a
        posterior distribution. \autoref{fig:pal5-results-full} and
        \autoref{fig:pal5-results-panels} show the membership likelihood
        evaluated at the $50$th percentile of this distribution, per star. In
        \autoref{fig:pal5-heatmap} we sample over the full distribution and
        smooth by a 0.1~deg (or mas/yr) kernel to produce a smooth
        representation of the stream. As a consequence of the sampling we see
        the model's stream solution extends beyond the $\approx \pm5\degree$
        shown previously.  Though the model certainty and individual membership
        likelihoods are low, the stream path matches closely with findings in
        deeper photometric catalogues, like \citet{Ibata+2017}.

        While \Gaia{} alone is not photometrically capable of observing the main
        sequence of \stream{Pal\,5}'s tidal tails, the modeling framework we
        present here is capable of operating on outer-joined datasets. This
        allows us to easily combine deep photometric data with kinematics from,
        e.g., \Gaia{}.  We leave that to a future work, though expect that a
        simultaneous model in photometry and astrometry would help identify the
        more diffuse parts of the stream, which become washed out in our
        astrometric fit due to the substantial number of background stars.
        Still, this section highlights the capability of our model when applied
        to a noisy field with a diffuse stream.



\section{Discussion} \label{sec:discussion}

    We now provide a discussion of the methods presented in the paper, focusing
    on limitations of our approach in \autoref{sec:limitations} and comparison
    to existing methods in \autoref{sub:comparison}. In
    \autoref{sub:future_surveys} we discuss prospects for applying our method to
    existing and future datasets.

    \subsection{Caveats and Limitations}\label{sec:limitations}

        We now highlight the limitations of our method, and areas for potential
        improvement.
    
        First, our usage of neural networks ensures that our method is
        expressive, and capable of describing a diverse range of stream
        morphologies all under one framework. At the same time, neural network
        training is computationally expensive, and takes a long time on a local
        device. Our fits to \stream{GD-1} and \stream{Pal\,5} presented in this
        work were performed on a local computer without GPU support, though
        training time takes on the order of a few hours per stream. This can be
        improved substantially with the use of GPUs.
    
        Second, our fits to \stream{GD-1} in \S\ref{sec:results_gd1} rely on a
        theoretical isochrone. While the stream does appear to be decently
        characterized by a theoretical isochrone, there is room for improvement.
        Especially considering that globular clusters do not always follow
        simple single-stellar population (see, e.g., \citealt{MiloneMarino2022}
        for a review of this subject), it is worth considering whether more
        data-driven approaches can be used to model the photometry of a stream
        without imposing tight priors on the stellar population. 
    
        Third, we have used a combination of analytic distributions and more
        flexible machine learning methods like normalizing flows to describe the
        distribution of background stars. However, the field of the Galaxy is
        complicated, in all coordinates. The analytic distributions in our
        modeling (specifically the $\phi_2$ background distributions) are
        clearly approximate. While the analytic distributions might be made more
        complex, the alternative is to substitute for non-parameteric flexible
        models. We have postponed this to future work, since the analytic
        backgrounds provide a decent description in the fields of \stream{GD-1}
        and \stream{Pal\,5}. However, this might not be the case for other
        streams. Constructing flexible background models will therefore provide
        further improvement over our method.

        Fourth, the number density of stars belonging to a stream is determined
        by several interesting physical processes. These include (e.g.)
        epicyclic overdensities, the orbital phase of the stream, and
        interactions with galactic substructure. Our model constrains the
        stellar number density along any given stream, and can therefore be used
        to infer the underlying mechanism(s) responsible for the stream’s
        density. However, all observational surveys are subject to selection
        effects. In order to determine the nature and reality of a stream’s
        density fluctuations, selection effects must be accounted for. For the
        case of \Gaia{}, there is ongoing work to model the survey’s selection
        function \citep{Cantat-Gaudin+2023}. Our model provides the first step,
        which is to characterize the observations.  With additional work, the
        selection function may be incorporated into our model and fits as a
        prior on the data.  Doing so will enable a more robust analysis of each
        stream's linear density, allowing us to make more definitive conclusions
        about its intrinsic structure. We defer this to future work
    
        Finally, our method is suited for characterizing streams rather than
        detecting them. Thus, some amount of supervised work is required to
        reduce the data down to a clean enough selection of stars so that our
        model can converge to a high likelihood solution. Crucially, the field
        must contain not only stream stars but also enough background stars to
        model that component well and distinguish stream from background. As a
        rule of thumb, it is best if the stream is at least a few percent of the
        total data, preferably even 10\%. Better background models permit this
        fraction to be pushed down further.  If the field is too noisy and
        control points cannot be reliably placed, our method might not be able
        to characterize the stream of interest.  This also applies to streams
        with bifurcated features like GD-1's spur component. To model these
        additional features we must add a separate mixture component. A more
        data-driven approach to characterize previously unknown or diffuse
        extensions to a stream would mark a substantial improvement over our
        current implementation. We leave this to future work.

    \subsection{Comparison to Existing Methods} \label{sub:comparison}

        There are several existing methods for the detection of stellar streams,
        and some existing work on detailed stream characterization. Here we
        briefly overview the existing methods, and highlight commonalities and
        differences between the results of our work and previous studies. 

        Currently, the matched filter is the most common method for detecting
        and characterizing streams (e.g., \citealt{Rockosi+2002, Grillmair2014,
        Bonaca+2012, Jethwa+2018, Shipp+2018}).  The method works by selecting a
        stellar population in the CMD (i.e., from a metal-poor isochrone) and
        searching for a stellar overdensity in the astrometric data. The
        approach has proved successful for stream detection, but to a lesser
        extent for characterization. From the matched filter approach alone,
        stream membership probabilities are not obtained and there is
        substantial contamination from background stars. While the approach can
        provide an excellent coarse-grained description of a given stream, it
        does not provide a density model in astrometry and photometry. 

        To a good approximation streams are expected to roughly trace a stellar
        orbit. The STREAMFINDER algorithm exploits this physical expectation
        \citep{STREAMFINDER}, by integrating orbits in a trial Galactic
        potential and identifying stellar overdensities that appear to coincide
        with segments of the trial orbits. While the method has identified many
        stream candidates, using these streams for potential reconstruction is
        circular in reasoning. That is, the detection and characterization of
        these streams is conditioned on a choice of the potential. The method we
        present here does not make assumptions about the gravitational
        potential, nor do we enforce physical assumptions about the dynamics of
        each stream. 

        This work represents one machine learning-based approach for stream
        characterization. Another method is presented in \citet{Shih+2022}, VIA
        MACHINAE, which uses conditional density estimation and a linear feature
        detector to discover streams and identify probable members. While the
        method is promising for stream discovery, its intended use is not for
        detailed characterization (as stated in \citealt{Shih+2022}). Our work
        has the opposite intent: the method presented is not designed to detect
        streams, rather, it is well-suited for characterizing streams that fall
        within some predefined field. Thus, our method can be used for
        characterizing the VIA MACHINAE candidates, and obtaining more robust
        membership probabilities. 

        In \citet{Patrick+2022} the stellar density of 13 tidal streams is
        modeled in on-sky positions and color-magnitude coordinates. The
        morphology of each stream is characterized by a series of splines, which
        are themselves a function of $\phi_1$. This method was originally
        developed in \citet{Erkal+2017}.  Importantly, \citet{Patrick+2022}
        performs a fit in $\phi_1-\phi_2$ and photometry, similar to our work.
        The photometric approach in \citet{Patrick+2022} relies on generating
        mock stellar populations in the CMD, selecting probable stream stars,
        modeling the $\phi_1-\phi_2$ distribution for those stars, and then
        refining the CMD fit based on the $\phi_1-\phi_2$ fit. This approach is
        therefore similar to a matched filter analysis. The model does not
        currently incorporate kinematic information, though the work could be
        extended. The improvement of our work is that we do not require an
        iterative fit to the CMD and astrometry: we handle both spaces through a
        joint likelihood without simulating mock-observations. Additionally, we
        can easily incorporate kinematic information and data with missing
        phase-space dimensions.

        We also highlight upcoming work from the CATS Collaboration 2023 (in
        prep.), which provides a spline-based characterization of streams
        implemented in the \package{Jax} library \citep{jax2018github}. The
        \package{Jax} backend ensures that fitting the model to a stream is very
        fast (on the order of minutes), and also has the capability of
        incorporating kinematic data with missing phase-space dimensions. The
        method does not currently model the color-magnitude distribution of a
        stream, but relies on a loose photometric selection. The photometric
        model in this work could be extended to the spline-based method,
        allowing for a joint astrometric and photometric model than can be
        rapidly fit to the stream of interest.

        In total, having multiple methods to characterize streams and their
        membership probabilities is advantageous. As the volume of tidal
        features continues to grow, it will become important to fit different
        models to the same stream, in order to evaluate whether certain features
        or density variations are real versus spurious artifacts from the
        adopted method. Considering the information encoded about perturbations
        in the small-scale density fluctuations along a given stream, a
        posterior distribution over models will help capture systematic errors
        in our description of streams and the field of perturbations that give
        rise to their density.  


    \subsection{Combining Information from Present and Future Surveys} \label{sub:future_surveys}

        Streams have been primarily discovered and characterized using
        photometric searches and matched filter methods (e.g.,
        \citealt{Rockosi+2002, Grillmair2014, Bonaca+2012, Jethwa+2018,
        Shipp+2018}).  \Gaia{} has enabled a more precise localization of
        streams, thanks to their  compact kinematic distribution relative to a
        field of background stars.  On the other hand, \Gaia{} is not
        photometrically deep ($G\lesssim 21$) and therefore is less sensitive to
        diffuse streams in the outer Milky Way halo.  Current and upcoming
        surveys --- e.g., DES \citep{Abbott+2018, Abbott+2021}, \textit{Euclid}
        \citep{Laureijs+2011}, Rubin \citep{Ivezic+2019}, \textit{Roman}
        \citep{Spergel+2013} --- do and will probe much deeper photometrically
        than \Gaia{} can detect.  From the more distant stellar halo
        \citep{Shipp+2018} to extragalactic systems
        \citep{Martinez-Delgado+2008, Demirjian+2020, Roman+2023}, these
        missions will reveal many new tidal features. Since \Gaia{} cannot be
        used for these streams, kinematic information will be rare, requiring
        dedicated followup observations.

        In this work we build mixture models (specifically MDNs) to characterize
        streams using all of the existing photometric and astrometric data for a
        given field. Looking to \autoref{fig:gd1-results-full} and
        \autoref{fig:pal5-results-full}, the proper motion components of the
        model are among the most information-rich and contribute significantly
        to the model. What then are the prospects for using this work with
        streams discovered with data from DES, \textit{Euclid}, Rubin, etc?
        There are two categories of streams these surveys will observe. The
        first is streams that partially appear in \Gaia{} but are deeper in
        these surveys; \stream{Pal\,5} is an example, with its main sequence
        below the \Gaia{} observation limit. Our method permits missing
        phase-space dimensions, therefore allowing ``outer''-joined catalogues
        mixing \Gaia{} and photometrically deeper surveys. In this scenario the
        stream track in proper motions and parallaxes (and possibly radial
        velocities) communicate and are mutually constrained with the
        photometric tracks. The second category is streams entirely missing from
        \Gaia{}, where the data is on-sky positions ($\phi_1$, $\phi_2$),
        photometry ($\mbs{m}$), and potentially metallicities. Even in this
        limited data regime,  our stream characterization framework is
        applicable. The  normalizing flow background model with a tightly
        constrained stream isochrone model can be flexibly applied to any
        stream, allowing for a characterization of the tidal feature insofar as
        the available data allows.  
        
        The ability to simultaneously model the on-sky and photometric
        distribution of a stream makes this approach a quantitative and Bayesian
        form of a matched filter analysis. For this reason, our method is
        readily applicable to the wealth of data expected from upcoming surveys,
        including both streams discovered in the Milky Way and extragalactic
        tidal features. While the present work is concerned with the former
        case, similar methods to the approach developed here may be applied to
        extragalactic tidal features, 1000s of which are expected to be
        discovered with Rubin \citep{Ivezic+2019} and \textit{Roman}
        \citep{Spergel+2013}. With a density model of an extragalactic tidal
        feature, the gravitational potential of external galaxies can be
        constrained (e.g., \citealt{Fardal+2013, Pearson+2022, Nibauer+2023}).
        We leave an extension of our method to extragalactic streams to future
        work.


\section{Summary and Conclusion} \label{sec:conclusions}

    We have constructed and demonstrated a new method for characterizing stellar
    streams, given all of the astrometric and photometric data available. We
    model a stream in astrometric coordinates using mixture density networks,
    allowing for a flexible representation of diverse stream morphologies under
    a single modeling framework. Additionally, our astrometric fit does not rely
    on assumptions about the gravitational potential. Our simultaneous
    photometric model treats each stream as a single-stellar population,
    generated from a theoretical isochrone. The distance modulus along the
    stream is itself a function of $\phi_1$, allowing us to tie photometric
    information to parallax measurements along the stream. This tie enables a
    reconstruction of stellar stream distance tracks that are compatible with
    photometry and astrometry by construction. A joint model in astrometric and
    photometric coordinates also naturally allows for more spatially diffuse
    regions of a stream to be detected. Additionally, our background model
    relies on a combination of analytic densities and normalizing flows, each of
    which are modular in \package{StreamMapper} and readily adaptable to a given
    field. 

    The combination of a flexible stream and background model allows us to
    obtain robust stream-membership probabilities for each star in a given
    field. This enables the generation of statistical samples for stream stars,
    informed by astrometry and photometry. We apply the method to the stream
    \stream{GD-1} and the tidal tails of \stream{Palomar 5}, generating
    membership catalogs for both streams that are publicly available at
    \href{https://zenodo.org/records/10211410}{zenodo:10211410}. Our
    characterization of \stream{GD-1} reveals a significant detection of the
    stream's main component as well as the ``spur" component, which represents a
    bifurcation from the main stream.  We also recover density variations and
    gaps along the stream that have been previously observed. Our fit to
    \stream{Pal\,5} demonstrates the performance of our model on a more diffuse
    stream with low signal-to-noise measurements.  The inclusion of deeper
    photometric data for \stream{Pal\,5} will enable a more extended view of the
    stream.

    Our method represents a new avenue to characterize the growing census of
    stellar streams discovered in the Milky Way. While most streams have been
    roughly characterized with matched filter algorithms, a statistical and
    homogeneous determination of stream-membership probabilities for each star
    has not been performed or made publicly available. A homogeneous catalog of
    stellar streams and stellar membership probabilities will therefore
    represent a substantial step forward for modeling the Galaxy, since both the
    global track of a stream and its small-scale density distribution are
    sensitive to the Milky Way's gravitational potential. 
    
    Especially when considering the small-scale perturbation history of a stream
    predicted by $\Lambda\rm{CDM}$, the inclusion or exclusion of stream stars
    can make a substantial difference in our inference of the perturbers.
    Constructing catalogs with membership probabilities as we do in this work
    provides a crucial step forward to making our inference of, e.g., dark
    matter subhalos in the Galaxy more robust. Equipped with membership
    probabilities, one is finally able to propagate the uncertainty in our
    determination of the ``on-stream data" through to the posterior distribution
    of model parameters (e.g., those characterizing the subhalo mass function).
    Thus, our works provides a new technique to generate samples of stream stars
    using all of the existing kinematic and photometric data, from which models
    for the Milky Way and its potential can be constrained in a statistically
    sound manner.


\newpage
\section{acknowledgements}

    NS acknowledges support from the Natural Sciences and Engineering Research
    Council of Canada (NSERC) - Canadian Graduate Scholarships Doctorate Program
    [funding reference number 547219 - 2020].  NS received partial support from
    NSERC (funding reference number RGPIN-2020-04712) and from an Ontario Early
    Researcher Award (ER16-12-061; PI Bovy).

    JN is supported by a National Science Foundation Graduate Research
    Fellowship, Grant No. DGE-2039656. Any opinions, findings, and conclusions
    or recommendations expressed in this material are those of the author(s) and
    do not necessarily reflect the views of the National Science Foundation.
    
    Both NS and JN would like to thank the
    \href{https://stellarstreams.org/streams22/}{Community Atlas of Tidal
    Streams 2022} conference for starting the discussions that led to this work.
    We thank Joshua Speagle for conversations and suggestions that were crucial
    to the development of the photometric model. We also thank Shirley Ho and
    Miles Cranmer for insightful discussions and comments.

    This work has made use of data from the European Space Agency (ESA) mission
    \Gaia{} (\url{https://www.cosmos.esa.int/gaia}), processed by the \Gaia{}
    Data Processing and Analysis Consortium (DPAC,
    \url{https://www.cosmos.esa.int/web/gaia/dpac/consortium}). Funding for the
    DPAC has been provided by national institutions, in particular the
    institutions participating in the \Gaia{} Multilateral Agreement.

    The data availability statement is modified from one provided to
    ShowYourWork by Mathieu Renzo.

    \paragraph{Software (alphabetical)}

        \package{asdf} \citep{Greenfield+2015}, %
        \package{astropy} \citep{Astropy2013, Astropy2018, Astropy2022}, %
        \package{astroquery} \citep{Astroquery2019}, %
        \package{brutus} \citep{brutus}, %
        \package{Matplotlib} \citep{Hunter2007}, %
        \package{NumPy} \citep{Harris+2020}, %
        \package{Pytorch} \citep{Pytorch2019}, %
        \package{SciPy} \citep{Scipy2020}, %
        \package{ShowYourWork} \citep{Luger+2021}, %
        \package{zuko} \citep{zuko}. %

\newpage
\section*{Data Availability} \label{sec:data_availability}

    This study was carried out using the reproducibility software
    \href{https://github.com/showyourwork/showyourwork}{\showyourwork}
    \citep{Luger+2021}, which uses continuous integration to programmatically
    download the data, perform the analyses, create the figures, and compile the
    manuscript. Each figure caption contains two links: one to the dataset used
    in the corresponding figure, and the other to the script used to make the
    figure. The datasets are stored at
    \href{https://zenodo.org/records/10211410}{zenodo:10211410}. The git
    repository associated with this study is publicly available at
    \href{https://github.com/nstarman/stellar_stream_density_ml_paper}{nstarman/stellar\_stream\_density\_ml\_paper}.
    The open-source code base developed for this project is
    \href{https://github.com/GalOrrery/stream_mapper-pytorch}{\package{StreamMapper}}.


\newpage
\bibliography{bib}

\appendix

\section{Probability Distributions} \label{app:distributions}

    A variety of probability distributions are used when modeling the stream and
    Galactic background. This appendix presents relevant mathematical details of
    these distributions.  In particular, observational errors are modeled as
    Gaussian distributions, which requires modifying each distribution to
    account for the Gaussian noise.

    \vspace{10pt}

    Let $x \sim X$ be distributed as some specified distribution and $\delta
    \sim \Delta \equiv \mcal{N}(0, \sigma_*)$ as an independent, centered
    Gaussian.  An observation $Y$ is distributed as
    \begin{equation}
        Z = X + \Delta,
    \end{equation}
    Which is a convolution of the PDFs.
    \begin{align} \label{eq:general_convolution}
        \pdf_Z(x; \mbs{\theta}, \sigma)
            & \equiv (\pdf_X(x; \mbs{\theta}) * \mcal{N}(x; 0, \sigma)) \\
            &= \int_{-\infty}^{\infty} \rm{d}\tau \, \pdf(x; \mbs{\theta})\mcal{N}(x-\tau; 0, \sigma)
    \end{align}

    In the following subsections we work through the distributions
    $\pdf(x,\mbs{\theta})$ and the convolutions with $\mcal{N}(x,0,\sigobs)$.

    \vspace{5pt}
    \subsection{Flat Distribution} \label{sub:flat_distribution}

        The simplest distribution is that of the univariate uniform
        distribution.  For a domain $x \in [a, b]$ the PDF is given by
        \begin{equation} \label{eq:pdf_flat_univariate}
            P_{\mcal{U}}(x|a,b) = \begin{cases}
                (b-a)^{-1} & a < x \leq b \\
                0 & \text{else}
            \end{cases}
        \end{equation}
        where $a,b$ are the bounds.

        In practice, the bounds $a,b$ are the bounds of the observation window,
        creating a subset of a larger field.  Suppose the larger field is
        uniformly distributed in a region larger than the bounds $a, b$, ie.
        $P_{\mcal{U}}(x|\alpha,\beta)$.  Let each observed datum $x_n$ in the
        field have Gaussian error $\sigobs_n$, then the PDF of the
        uniform-Gaussian-convolved distribution is given by:

        \begin{equation} \label{eq:flat_distribution}
            \pdf_{(\mcal{U}*\mcal{N})}(x; \alpha, \beta, \sigobs) = \begin{cases}
                \pdf_\mcal{U}(x; \alpha, \beta) \left( \cdf_{\mcal{N}}(x; \alpha, \sigobs) - \cdf_{\mcal{N}}(x; \beta, \sigobs) \right) & \alpha \leq x \leq \beta \\
                0 & \text{else}
            \end{cases}
        \end{equation}

        We then truncate this distribution to the observed field $(a, b]$
        \begin{equation} \label{eq:pdf_flat_univariate_convolved_error_full}
            \pdf_{(\mcal{U}*\mcal{N})}(x | a < X \leq b; \alpha, \beta, \sigobs)
        \end{equation}
        in the normal fashion: setting $\pdf_{(\mcal{U}*\mcal{N})} = 0$ when $x
        < a, x > b$ and normalizing $\pdf_{(\mcal{U}*\mcal{N})}$ within $(a,b]$.
        The truncated distribution is tractable but unwieldy. The distribution
        can be simplified enormously if two conditions hold: first, $\alpha \ll
        a, b \ll \beta$, the observed field is a small subset of the full field;
        and second, $\sigobs \ll |\alpha - a|$ and $\sigobs \ll |\beta - b|$,
        the errors are smaller than the observation window size compared to the
        full field.  In this case, the distribution reduces back to the uniform
        distribution!

        Therefore, for all cases considered in this work
        \eqref{eq:pdf_flat_univariate_convolved_error_full} becomes:

        \begin{equation}\label{eq:pdf_flat_univariate_convolved_error}
            \pdf_{(\mcal{U}*\mcal{N})}(x | a < X \leq b; \alpha, \beta, \sigobs) \cong P_{\mcal{U}}(x|a,b)
        \end{equation}

    \vspace{10pt}
    \subsection{Truncated Exponential Distribution} \label{app:sub:exponential_distribution}
    
        The univariate exponential distribution is:
        \begin{equation} \label{eq:pdf_exp_univariate}
            \pdf_{\mcal{E}}(x; \lambda) = \begin{cases}
                m \Exp{-\lambda x} & x \in [0, \infty) \\
                0 & x < 0
            \end{cases}
        \end{equation}
    
        The support of this distribution, $[0, \infty)$, rarely matches the
        support of the data. We generalize to the unit-normalized, truncated,
        univariate exponential distribution in the domain $(a, b]$:
        \begin{equation} \label{eq:pdf_truncexp_univariate}
            \pdf_{\mcal{E}_T}(x; \lambda, a, b) = \begin{cases}
                \frac{\lambda \Exp{-\lambda \, (x - a))}}{1 - \Exp{-\lambda(b - a)}} & a < x \leq b \\
                0 & \text{else},
            \end{cases}
        \end{equation}
        for bounds $x \in (a,b]$. However, as $\lambda \rightarrow 0$, which
        describes the flat distribution (\autoref{sub:flat_distribution}), this
        functional form can be numerically unstable. It is therefore practical
        to define $\pdf_{\mcal{E}}(x; |\lambda| < \epsilon, a, b) \cong
        P_{\mcal{U}}(x; a,b)$.

        Let each observed datum $x_n$ in the field have Gaussian error
        $\sigobs_n$, then the PDF of the Exponential-convolved-with-a-Gaussian
        distribution is given by:

        \begin{equation}
            \pdf_{\mcal{E}\cdot\mcal{N}}(x;m,a,\sigma_*) = \frac{m  e^{\frac{1}{2} \lambda  \left(2 b+\lambda  \sigma _*^2-2 x\right)}}{e^{\lambda  (b-a)}-1} = \pdf_{\mcal{E}}(x;\lambda,a) \ \Exp{\frac{1}{2}\lambda^2 \sigma_*^2},
        \end{equation}
        where the function is no longer truncated, since a Gaussian is not
        compact.

        We impose the same observation field $(a, b]$ and truncate
        $\pdf_{\mcal{E}\cdot\mcal{N}}(x | a < X \leq b;m,a,\sigma_*)$.  We note
        the normalization, which is the CDF evaluated at $a, b$, cancels the
        factor $\Exp{\frac{1}{2}\lambda^2 \sigma_*^2}$. Following thtough, the
        truncated convolved distribution is identical to the original
        distribution.

        \begin{equation}
            \pdf_{(\mcal{E}\cdot\mcal{N})_T}x | a < X \leq b;m,a,\sigma_*) \equiv \pdf_{\mcal{E}}(x; \lambda, a, b)
        \end{equation}

    \vspace{10pt}
    \subsection{Normal Distribution} \label{app:sub:normal_distribution}

        The properties of the Normal distribution are well known. For
        completeness we state

        \begin{equation}
            P_X(x; \mu, \sigma) = \mcal{N}(x; \mu, \sigma).
        \end{equation}

        With the inclusion of observational errors the distribution becomes

        \begin{equation}
            P_{\mcal{N}\cdot\mcal{N}}(x; \mu, \sigma, \sigobs) = \mcal{N}(x; \mu, \sqrt{\sigma^2 + \sigobs^2}).
        \end{equation}

        The Normal distribution is not compact, with non-zero value over all
        $x$. In an observational window $(a,b]$ the PDF is truncated, giving

        \begin{equation}
            P_{\mcal{N}_T}(x; \mu, \sigma, a, b) = \begin{cases}
                \frac{\mcal{N}(x; \mu, \sigma)}{\Phi_\mcal{N}(b; \mu, \sigma) - \Phi_\mcal{N}(a; \mu, \sigma)} & a < x \leq b \\
                0 & \text{else}
            \end{cases}
        \end{equation}

        Note that the observation-window-truncation is performed on the full
        distribution, which includes the observational errors, if present.
        Therefore, let $\tilde{\sigma}^2 \equiv \sigma^2+\sigma_*^2$

        \begin{equation}
            P_{(\mcal{N}\cdot\mcal{N})_T}(x; \mu, \tilde{\sigma}, a, b) = P_{\mcal{N}_T}(x; \mu, \tilde{\sigma}, a, b)
        \end{equation}

\newpage
\section{Pairing Neural Networks with Priors} \label{app:nns_and_priors}

    In modeling, the choices of priors are important, not only for the
    information they adds to a model, but also for how optimizers explore
    parameter spaces. Simple optimizers, like MCMC, propose a parameter vector
    then compute its likelihood given the data \citep{HoggForeman-Mackey2018}.
    Priors with compact support  $\theta \sim \mcal{D}((a, b])$, e.g.
    $\mcal{U}(a, b)$, may be very dangerous as the optimizer can be initialized
    in a region with zero probability and has no means but blind luck to find
    the region of support -- the region with non-zero likelihood. More
    sophisticated optimizers, like gradient-based descent \citep{cauchy_2009},
    are susceptible to a similar initialization problem, allowing
    zero-probability regions, but requiring at least a non-zero gradient to make
    informed updates to the optimizers state.  Outside the support the gradient
    can be zero.  With MDNs, model parameters are neural network outputs. Unlike
    for conventional models where an informed initial parameter state may be
    chosen by hand, it is impractical and undesirable to initialize by hand a
    neural network's output by setting all its internal weights. One might
    imagine accomplishing a desired initial output by pre-training the network
    with a toy model that forces convergence on a set output vector.  However,
    neural networks are highly non-linear in their response to internal state
    changes (a desirable property). It is quite possible that when switching
    from the toy to real model's loss function the pre-trained network is in a
    local probability island from which the global optimum is unreachable.  
    Therefore, with the MDN neural networks, initialization is mostly
    for checking the dependence of the results on the initial state --
    though even this dependence may instead be done using parameter
    dropout. Rather than focusing on initializing to a particular
    parameter vector, for neural networks it is instead important that
    the initial state, whatever it is, has non-zero probability (or
    non-zero gradient, depending on the optimizer).  Neural networks
    have some intrinsic phase-space range that depends on the
    architecture of the network. Therefore the network architecture must
    be chosen such that the outputs match well with the priors. For
    example, with a uniform prior $\theta \sim \mcal{U}(a,b)$, the
    network should be restricted such that $\theta \in (a, b]$, e.g. by
    ending the network's $\theta$ channel with a scaled logistic
    function.

\end{document}